
\magnification\magstep 1
\vsize=24 true cm
\hsize=16 true cm
\baselineskip=18pt
\parindent=1 true cm
\parskip=0.6 true cm
\def\auj{\number\day\space\if case\month\or
    janvier\or f\' evrier\or mars\or avril
    \or mai\or juin\orjuillet\or ao\^ut
    \or septembre\or octobre\or novembre            
    \or d\' ecembre\fi\space\number\year}		


\def\boxit#1#2{\setbox1=\hbox{\kern#1{#2}\kern#1}%
\dimen1=\ht1 \advance\dimen1 by #1 \dimen2=\dp1
\advance\dimen2 by #1
\setbox1=\hbox{\vrule height\dimen1 depth\dimen2\box1\vrule}%
\setbox1=\vbox{\hrule\box1\hrule}%
\advance\dimen1 by .4pt \ht1=\dimen1
\advance\dimen2 by .4pt \dp2=\dimen2 \box1\relax}

\def\ok{\displaystyle}
\def\ecart{\noalign{\medskip}}
\def\gsim{{{}_>\atop{}^\sim}}
\def\psim{{{}_<\atop{}^\sim}}
\def\operarrow{\mathop{\rightarrow}}

\def\opsim{\mathop{\sim}}
\def\opsimeq{\mathop{\simeq}}
\def\oppropto{\mathop{\propto}}
\def\${{1\over 2}}
\def\t{\theta}
\def\T{(\theta)}

\def\page0{\font\bfmag=cmbx8 scaled                
\magstep2\parskip=0.2 true cm
\baselineskip=14pt\nopagenumbers\vglue 1cm}


\def\rf#1,{$^{[#1]}$}                               

\def\2aut#1#2, #3#4,{#1{$^#2$} and #3{$^#4$}}
\def\1aut#1{\vglue 1cm\centerline{#1}\vglue 0,5cm}


\newtoks\auteur                                        
\auteur={\hfil}                                        
\font\eightsl=cmsl8                                    
\headline{    \hfill\eightsl\the\auteur}


\def\labo{\begingroup\noindent
               \sl Division de Physique Th\'eorique\footnote{*}
               {Unit\' e de Recherche des Universit\' es
                Paris 11 et Paris 6 Associ\' ee au CNRS},     
               Institut de Physique Nucl\' eaire,
               91406 Orsay Cedex and LPTPE,
               Universit\'e P. \& M. Curie,
               4 Place Jussieu 75252 Paris Cedex 05
               (France)\par\endgroup}


\def\abs#1{\vglue 2cm{\noindent\bf ABSTRACT ---            
            }\begingroup\narrower                          
           \baselineskip=12pt{#1}\par\endgroup}

\def\ipno#1,#2,#3,{\noindent IPNO/TH #1-#2\hfill #3 #1\eject}   

\pageno=1

\def\Ref{\noindent{\bf REFERENCES}\parindent=1.5 cm}
\def\ref#1{\item{\hbox to \parindent{\enskip \bf [#1]\hfill}}}


\def\Fig{\noindent{\bf FIGURE CAPTIONS}\parindent=1.5 cm}
\def\fig#1{\item{\hbox to \parindent{\enskip \bf Fig. #1\hfill}}}

\vglue 2cm
\centerline{\bf{LOCALIZATION PROPERTIES}}
\centerline{\bf{IN ONE DIMENSIONAL DISORDERED}}
\centerline{\bf{SUPERSYMMETRIC QUANTUM MECHANICS}}
\vglue 1cm
\centerline{Alain COMTET, Jean DESBOIS and C\'ecile MONTHUS}
\vglue 1cm
\labo

\abs {A model of localization based on the Witten Hamiltonian of supersymmetric
quantum mechanics is considered. The case where the superpotential $\phi(x)$ is
a random telegraph process is solved exactly. Both the localization length and
the density of states are obtained analytically. A detailed study of the low
energy behaviour is presented. Analytical and numerical results are presented
in the case where the intervals over which $\phi(x)$ is kept constant are
distributed according to a broad distribution. Various applications of this
model are considered.}
\vfill
\ipno 94 ,33 ,APRIL,

\item{\hbox to\parindent{\enskip \bf {I -} \hfill}}
{\bf INTRODUCTION}

Since the pionnering work of Anderson [1], the localization problem in one
dimension
 has been
investigated extensively and is by now pretty well understood [2]. However,
despite
the large number of rigorous results, there are very few solvable models in the
continuum for which one can compute exactly the density of states and the
localization length. The purpose of this work is to present some new results
for such a model which is defined by the one dimensional Schr\"odinger
Hamiltonian
$$
H=-{d^2\over dx^2}+\phi^2(x)+\phi'(x)
\eqno(1.1)
$$
This Hamiltonian, which was introduced by E. Witten as a toy model of
supersymmetric quantum mechanics [3], has stimulated a number of interesting
developments in the context of quantum mechanics. In particular it has provided
exact solutions of the Schr\"odinger equation for a class of so called shape
invariant potential [4]. It has also inspired a new method of semi-classical
quantization [5,6]. In the context of disordered systems, i.e. when the
 superpotential
$\phi(x)$ is considered as random, one of the most interesting features of this
model comes
from its relation to the problem of classical diffusion of a particle in a one
dimensional random medium. This correspondence [7], which has been exploited in
great detail in [8], is based on the observation that the imaginary time
Schr\"odinger equation
$$
{\partial \psi\over \partial t}+H\psi=0
\eqno(1.2)
$$
can be cast into the Fokker-Planck equation
$$
{\partial P\over \partial t}={\partial \over \partial x}\left({\partial P\over
\partial x}
-2\phi P\right)
\eqno(1.3)
$$
through the transformation
$$
\psi(x,t)=e^{-\int^xdy\ \phi(y)}\ P(x,t)
$$
Alternatively, the diffusion process is described by the Langevin equation
$$
\dot x(t)=2\phi(x)+\eta(t)
\eqno(1.4)
$$
where $\eta(t)$ is a gaussian white noise such that
$$
\overline{\eta(t)\eta(t')}=2\delta(t-t')
\eqno(1.5)
$$
This correspondence enables one to express the transition probability $P(x \ t\
\vert
y \  0)$ in terms of the eigenstates of $H$. In the disordered case, namely
when
$\phi(x)$ is random, the probability of returning to the starting site,
averaged over the disorder, becomes
$$
<P(x\ t\ \vert\ x\ 0)>=\int_0^\infty e^{-Et}\rho(E) \ dE
\eqno(1.6)
$$
The long time behaviour of $P$ is thus related to the low energy behaviour
of the density of states $\rho(E)$.

In the case of a white noise
$$\eqalign{
&<\phi(x)>=\mu\sigma\cr
&<\phi(x)\phi(x')>-<\phi^2(x)>=\sigma\delta(x-x')\cr
}
\eqno(1.7)
$$
both the localization length and the density of states have been computed
exactly [8,9]
and display unusual behaviours at low energy. Denoting by N(E) the integrated
density
of states, one obtains
\noindent
$\displaystyle {N(E) \oppropto_{E\to 0}}{1\over \ln^2E}$\quad for $\mu=0$ \ ,\
and
$\displaystyle{N(E)\oppropto_{E\to 0}}E^\mu$ \quad for $\mu>0$.

This implies that $<P(xt\ \vert\ x\ 0)>$ will present an algebraic tail
at large time which reflects the anomalous behaviour of the position $x(t)$.
Since the anomalous behaviour of the diffusion can be predicted qualitatively
[10],
the above correspondence can be used to understand physically the
origin of the anomalous behaviour of the localization problem (see also
[11,12]).

However a direct interpretation of these results without appealing to such a
correspondence has until now been missing. One of the purposes of this paper is
to fill
this gap. In our analysis, a crucial role will be played by the zero energy
states
$\psi_0(x)=\exp\int^x\phi(t)dt$ and
$\psi_1(x)=\psi_0(x)\int^x{dy\over\psi_0^2(x)}$ which are exactly known for any
realization of the disorder. This very unusual feature in the context of
disordered systems can of course be traced back to the supersymmetric structure
of the Hamiltonian.

In order to display this structure it is convenient to introduce the pair of
Hamiltonians
$$
H_\pm=-{d^2\over dx^2}+\phi^2(x)\pm\phi'
\eqno(1.8)
$$
They can be rewritten in the factorized form
$$
 H_+ = Q^{\dagger} Q \quad \hbox{and} \quad  H_- = Q Q^{\dagger} \quad \hbox{,
where}
 \quad Q= -{d \over dx} + \phi(x)
\eqno(1.9)
$$
This implies $H_+$ and $H_-$ have the same spectrum for $E>0$ [3]. In the
presence of
disorder they are characterized by the same Lyapunov exponent and density of
states. We will see that the possibility to treat them on the same footing will
significantly simplify our analysis.

In all this work, the disorder is modeled by assuming that the potential
$\{\phi(x)\}$ is described by an ensemble of rectangular barriers with
alternating heights $\phi_0$ and $\phi_1$ of random length. Most of our work
deals with the case where $\{\phi(x)\}$ is a random telegraph process [13]. The
lengths of the intervals over which $\phi(x)$ is constant are therefore
distributed according to an exponential law $p_i(\ell)= \theta(\ell) \
n_i\exp(-n_i\ell)$ for
$i=0,1$. Besides its mathematical interest, this form of disorder is
motivated by the fact that it can model certain quasi-one-dimensional
structures
with a piecewise constant order parameter [14]. Another advantage is that this
type
of disorder is less singular than a pure white noise process for which
arbitrary jumps of the potential can occur. An interesting extension of this
model corresponds to the case where the lengths of the barriers are chosen
according to a broad distribution which behaves for large $\ell$ as
$\ell^{-(1+\alpha)}$.

This paper is organized as follows. In part II we present a method to
compute the density of states which is an adaptation of the one discussed by
Benderskii and
Pastur [15]. We then apply this method to the case where the disorder is a
random
telegraph process. Exact analytical expressions of the density of states and
localization length are derived. This is followed by a careful analysis of the
limiting behaviour for $E\to 0$. A physical interpretation of these results is
given in part IV.
Numerical simulations are presented in part V. Then the case where $p(\ell)$
is a broad distribution is considered. Remarkably enough, both the density of
states and localization length can be computed exactly when $\alpha<1$ and
$\phi_1
\ = -\phi_0$. Finally,
in part VI we apply some of these results to the study of anomalous diffusion
and discuss various applications.

\vfill\eject
\item{\hbox to\parindent{\enskip \bf {II -} \hfill}}
{\bf THE PHASE FORMALISM FOR THE SUPERSYMMETRIC\break
HAMILTONIANS $H_\pm$}

The phase formalism is based on the following property of one-dimensional
Hamiltonians : the integrated density of states per unit length $N(E)$ and the
Lyapunov exponent $\gamma(E)$ are directly related to properties of the
solution $\psi_E$ of the corresponding Schr\"odinger equation
$H\psi_E=E\psi_E$ with a given logarithmic derivative at one point [16].
The number $N(E)$ of states of energy lower than $E$, per unit length, is
equal to the number of nodes of the wave function $\psi_E$ per unit length,
and the Lyapunov exponent $\gamma(E)$ measures the exponential growth rate of
the envelop of the wave function $\psi_E$.

For any  spatially homogenous disordered potential with short range
correlations,
$N(E)$ and $\gamma(E)$ are self-averaging quantities. This means that they are
the same for all realizations of the disorder with probability one.

A convenient way to implement the phase formalism for the Schr\"odinger
equations
$$
\left\{\matrix{
H_+u_E=Eu_E\hfill\cr
H_-v_E=Ev_E\hfill\cr
}\right.
\eqno(2.1)
$$
is to introduce "polar" variables
$$
\left\{\matrix{
u_E(x)=\rho_E(x) \cos\theta_E(x)\hfill\cr
v_E(x)=\rho_E(x) \sin\theta_E(x)\hfill\cr
}\right.
$$
The phase $\theta_E(x)$ contains all the informations about the oscillations
of the wave functions $(u_E,\ v_E)$ and the modulus $\rho_E(x)$ represents the
envelope of these wave functions. The new dynamical variables $\theta_E$ and
$\rho_E$ will determine the value of $N(E)$ and $\gamma(E)$\break
respectively.\break $N(E)$ is equal to the number of nodes of $u_E$ (or $v_E$)
per unit length and can thus be written directly in terms of $\theta_E(x)$
$$
N(E)={\displaystyle\lim_{L\to \infty   }}
\left(
{\theta_E(L)-\theta_E(0)\over\pi L}
\right)\ .
\eqno(2.2)
$$
Note that $[\theta_E(L)-\theta_E(0)]$ is the total phase accumulated on the
interval $[0,L]$ and is therefore not an angle defined modulo $2\pi$.
$\gamma(E)$ is the exponential growth rate of the envelope $\rho_E$ and reads
therefore
$$
\gamma (E) ={\displaystyle\lim_{L\to \infty   }} {1\over L}\ln
\left(
{\rho_E(L)\over\rho_E(0)}
\right)
\eqno(2.3)
$$
Now in order to write the dynamical equations for $\theta_E$ and $\rho_E$ we
have to separate the cases $E>0$ and $E<0$. From now on we will set $k=C{\vert
E\vert}$.

\noindent {\bf 1 --- case $E=+k^2$}

In this case, the factorized forms (1.9) allow to rewrite the system (2.1)
 as two coupled first order equations $Qu=kv$ and $Q^{\dagger}v=ku$,
 which explicitly read
$$\left\{\matrix{
\displaystyle {du_E\over dx}=\phi(x)u_E(x)-kv_E(x)\hfill\cr
\noalign{\medskip}
\displaystyle {dv_E\over dx}=-\phi(x)v_E(x)+ku_E(x)\hfill\cr
}\right.
\eqno(2.4)
$$
This gives the following dynamical equations for the "polar" variables
$$\left\{\matrix{
\displaystyle {d\theta_E\over dx}=k-\phi(x)\sin 2\theta_E(x)\hfill\cr
\noalign{\medskip}
\displaystyle {d\ln\rho_E\over dx}=\phi(x)\cos 2\theta_E(x)\hfill\cr
}\right.
\eqno(2.5)
$$
Note that the equation for $\theta_E$ does not involve $\rho_E$.

Let us now present the direct resolution of these equations in ($\theta_E,\
\rho_E$) for the special cases $E=0$ and $E\to \infty   $, before we turn to
the
general method to calculate $N(E)$ and $\gamma(E)$ for all $E>0$.
\medskip
\boxit{3pt}{a) case $E=0$}

For $k=0$, the dynamical equations (2.5) read
$$\left\{\matrix{
\displaystyle {d\theta\over dx}=-\phi(x)\sin 2\theta(x)\hfill\cr
\noalign{\medskip}
\displaystyle {d\ln\rho\over dx}=\phi(x)\cos 2\theta(x)\hfill\cr
}\right.
\eqno(2.6)
$$
The phase $\theta(x)$ cannot grow more than $\pi/2$ on the interval [0, L]
since, for any $\{\phi(x)\}$, the velocity ${d\theta\over dx}$ vanishes
whenever $\theta=0$ [modulo $\pi/2$]. Therefore (2.2) yields immediately
$N(0)=0$ in accordance with the positivity properties of $H_\pm$.

To get $\gamma(0)$, we can integrate the equation for $\theta(x)$
$$
\left\vert{\tan\theta(L)\over\tan\theta(0)}\right\vert=
e^{-2\int_0^L\phi(x)dx}
\eqno(2.7)
$$
and rewrite the equation for $\rho$ as
$$
{d\ln\rho\over dx}=-{d\theta\over dx}{\cos 2\theta(x)\over \sin 2\theta(x)}
\eqno(2.8)
$$
The integration gives
$$
\gamma(0)\equiv {\displaystyle\lim_{L\to \infty   }}
\left(
{\ln\rho(L)-\ln\rho(0)\over L}
\right)={\displaystyle\lim_{L\to \infty   }}{1\over 2L}\ln
\left\vert
{\sin 2\theta(0)\over \sin 2\theta(L)}
\right\vert
\eqno(2.9)
$$
We now use (2.7); in the limit $L \to \infty$, the expression
 ${1\over L}\int _0^L\phi(x)dx$ is
simply the mean value of  $\{\phi(x)\}$, that we will note $F_0\equiv <\phi>$
from now on. We obtain finally that the Lyapunov exponent at zero energy is
equal to the absolute value of the average $F_0$ of  $\{\phi(x)\}$
$$
\gamma(0)=\vert F_0\vert
\eqno(2.10)
$$

This simple result can also be recovered by starting from the two linearly
independent solutions of $H_+\psi=0$
$$\left\{\matrix{
\psi_0(x)=e^{+\int _0^x\phi(t)dt}\hfill\cr
\noalign{\medskip}
\psi_1(x)=\psi_0(x)\int   _0^x{dt\over\psi_0^2(t)}=
e^{\int   _0^x\phi(t)dt}\int   _0^xdy\ e^{-2\int   _0^y\phi(t)dt}\cr
}\right.
\eqno(2.11)
$$
The asymptotic behaviour for large $x$
$$
\int _0^x\phi(t)dt \opsim_{x\to \infty   } F_0 x [1+o(1)]
$$
allows to recover (2.10).
\medskip
\boxit{3pt}{b) Limit $E\to \infty   $}

At high energy, (2.5) can be approximated by
$\displaystyle {{d\theta\over dx}\opsim_{k\to \infty   }}k=C{E}$
and  $N(E)$ behaves therefore asymptotically as the free integrated
density of states
$\displaystyle {N(E)\opsim_{E\to \infty }{C{E}\over\pi}}$ .

At this order of approximation, the Lyapunov exponent vanishes
$$\eqalign{
\gamma(E)&=\lim_{L\to \infty }
<{1\over L}\int   _0^Ldx\phi(x)\cos[2\theta(x)]>\cr
\noalign{\medskip}
&\opsim_{E\to\ \infty   }\lim_{L\to \infty }<\phi>{1\over L}\int
_0^Ldx\cos(2kx)=0\cr
}$$
To find the asymptotic behaviour of $\gamma(E)$ we must calculate the next
order for $\theta(x)$
$$
\theta(x)\opsim_{k\to \infty }kx- \int _0^{x}\phi(x')\sin(2kx')dx'
\eqno(2.12)
$$
This allows to relate the high energy behaviour of $\gamma(E)$ to the two
point correlation function of the process $\{\phi(x)\}$ [16]
$$
\gamma(E)\opsim_{E\to \infty   }
{1\over 2}\int _0^{\infty}  dx \ \cos(2kx) \big[ <\phi(x)\phi(0)> - <\phi>^2
\big]
\eqno(2.13)
$$
This Fourier transform relation shows that the high energy decay of the
Lyapunov exponent
is directly related to the regularity of the random process $\{\phi(x)\}$.

For the white noise case $<\phi(x)\phi(x')>=<\phi>^2+\sigma\delta(x-x')$ we
recover the result [8]
$$
{\displaystyle{}^\gamma}{}_{WN}(E)\opsim_{E\to\ \infty   }{\sigma\over 2}
\eqno(2.14)
$$
This behaviour at high energy is rather pathological and very particular to
white noise. Indeed for an exponentially correlated noise
$$
<\phi(x)\phi(x')>=<\phi>^2+{\lambda\sigma\over 2}e^{-\lambda\vert x-x'\vert}
\eqno(2.15)
$$
which can be viewed as a regularization of the white noise case as long as
$\lambda$ is finite, we obtain a vanishing Lyapunov exponent
$$
\gamma (E) \opsim_{E\to \infty   } \left({\sigma\lambda^2\over 8}\right)
\ {1\over E}
\eqno(2.16)
$$

\medskip
\boxit{3pt}{c) General method to calculate $N(E)$ and $\gamma(E)$ for all
$E>0$}

We have seen that the dynamical equations (2.5) could be integrated directly
for the particular values $E=0$ and $E\to \infty   $ in order to get $N(E)$ and
$\gamma(E)$. This is obviously not possible in general. We therefore need a
more powerful method [15]. We can use the identity
$\displaystyle \delta[f(x)]=\sum_{x_i\  \hbox{\sevenrm zero of f}}
{\delta(x-x_i)\over\vert f'(x_i)\vert}$
 to rewrite the number
${\cal N}_L{(E)}$ of nodes of $\cos\theta_E(x)$ in the interval [0, L]
$$\eqalignno{
{\cal N}_L{(E)}&=\int   _0^L dx\sum_{x_i\ \hbox{\sevenrm
solution}\atop\hbox{\sevenrm
of} \cos\theta(x_i)=O}
\delta(x-x_i)=k\int   _0^Ldx\ \delta[\cos\theta(x)]\cr
&=k\int   _0^L dx \sum_{n\in Z}\delta[\theta(x)-{\pi\over 2}-n\pi]&(2.17)\cr
}$$
The integrated density of states $N(E)$ then reads [15]
$$
N(E)=\lim_{L\to \infty   }
\left(
{\cal N}_L{(E)}\over L\right)=kP_{eq}\left(\pi\over 2\right)
\eqno(2.18)
$$
where $P_{eq}(\theta)$ is the stationary distribution of the reduced phase
$\theta$ defined modulo $\pi$.

Similarly the Lyapunov exponent can be rewritten as [17]
$$
\gamma(E)=\lim_{L\to \infty   } {1\over L}\int_0^Ldx\ \phi(x)\cos 2\theta(x)=
<\phi(x)\cos2\theta(x)>
\eqno(2.19)
$$
where the brackets $<\cdots>$ denote the mean value taken over the joint
probability of the process $\phi$ and of the reduced phase $\theta$. We will
use these results in  section III to calculate explicitly $N(E)$ and
$\gamma(E)$ for a specific model of disorder.

\noindent {\bf 2 --- case $E=-k^2$}

We have already seen that $N(0)=0$ so that $N(E)$  vanishes identically for
$E\in ]-\infty ,\ 0]$. It is nevertheless interesting to study the Lyapunov
exponent
$\gamma(E)$ in this unphysical region since it is relevant for the problem of
classical diffusion in the random force field $\{\phi(x)\}$.

For $E<0$, the factorized forms (1.9) allow to rewrite the system (2.1)
 as two coupled first order equations $Qu=kv$ and $Q^{\dagger}v= -ku$,
 which explicitely read
$$\left\{\matrix{
\displaystyle {du_E\over dx}=\phi(x)u_E(x)-kv_E(x)\hfill\cr
\noalign{\medskip}
\displaystyle {dv_E\over dx}=-\phi(x)v_E(x)-ku_E(x)\hfill\cr
}\right.
\eqno(2.20)
$$
where $k=C{-E}$.

The dynamical equations for the polar coordinates then read
$$\left\{\matrix{
\displaystyle {d\theta_E\over dx}=-k\cos 2\theta_E(x)-\phi(x)\sin
2\theta_E(x)\hfill\cr
\noalign{\medskip}
\displaystyle {d\ln\rho_E\over dx}=-k\sin 2\theta_E(x)+\phi(x)\cos
2\theta_E(x)\hfill\cr
}\right.
\eqno(2.21)
$$
Note the differences with (2.4) and (2.5) respectively.

We recover immediately that $N(E)=0$ for $E<0$ by noting that the accumulated
phase $\theta(x)$ will be trapped in an interval of length $\leq {\pi\over
2}$. Indeed for any $\{\phi(x)\}$ the velocity
$d\theta\over dx$ is
positive
whenever $\theta={\pi\over 2}$ [modulo $\pi$], and
negative whenever
$\theta=0$ [modulo $\pi$].

Still we need to study the stationary distribution $P_{eq}(\theta)$ in order
to calculate $\gamma(E)$ along the lines explained in II 1 c).
\parindent= 1.5 true cm
\item{\hbox to\parindent{\enskip \bf {III -} \hfill}}
{\bf APPLICATION TO A MODEL OF RECTANGULAR BARRIERS OF RANDOM LENGTHS}
\parindent= 1 true cm

\noindent {\bf 1 --- Description of the model}

Let us now consider a model [15] where $\phi(x)$ takes alternatively two values
$\phi_0$ and $\phi_1$ on intervals whose lengths are positive independent
random variables (Fig.1), distributed according to the following probability
densities
respectively
$$\left\{\matrix{
f_0(l)= \theta(l) \ n_0 \ e^{-n_0l}\hfill\cr
\noalign{\medskip}
f_1(h)= \theta(h) \ n_1 \ e^{-n_1h}\hfill\cr
}\right.
\eqno(3.1)
$$
This choice for $f_0$ and $f_1$ is in fact the only one that makes the process
$\{\phi(x)\}$ Markovian. This property enables us to write differential
equations for the probability $p_0(x)$ to have $\phi(x)=\phi_0$ and the
probability $p_1(x)=1-p_0(x)$ to have $\phi(x)=\phi_1$.
$$\left\{\matrix{
\ok {\partial  p_0\over \partial x}=-n_0p_0+n_1p_1=n_1-(n_0+n_1)p_0\hfill\cr
\noalign{\medskip}
 \ok {\partial p_1\over \partial x}=-n_1p_1+n_0p_0=n_0-(n_0+n_1)p_1\hfill\cr
}\right.
\eqno(3.2)
$$
The corresponding stationary solutions are simply
$$\left\{\matrix{
\ok \lim_{x\to \infty }p_0(x)={n_1\over n_0+n_1}={<l>\over <l>+<h>}\hfill\cr
\noalign{\medskip}
\ok \lim_{x\to \infty }p_1(x)={n_0\over n_0+n_1}={<h>\over <l>+<h>}\hfill\cr
}\right.
\eqno(3.3)
$$
where
$<l>=\int    _0^\infty    dl\ lf_0(l)= \ok {1\over n_0}$ is the mean length of
intervals
$\{\phi(x)=\phi_0\}$\break
and
$<h>=\int  _0^\infty    dh\ hf_1(h)= \ok {1\over n_1}$ is the mean length of
intervals
$\{\phi(x)=\phi_1\}$.

The mean value $F_0$ and the two point correlation function $G(x)$ of the
process $\{\phi(x)\}$ read
$$\left\{\matrix{
F_0\equiv <\phi>=\phi_0 \ok{n_1\over n_0+n_1}+\phi_1{n_0\over n_0+n_1}\hfill\cr
\noalign{\medskip}
G(x)\equiv <\phi(x)\phi(0)>-<\phi>^2= \ok{n_0n_1\over(n_0+n_1)^2}
(\phi_1-\phi_0)^2e^{-(n_0+n_1)\vert x\vert}\hfill\cr
}\right.
\eqno(3.4)
$$
The case $\phi_1=-\phi_0$ where
$$\left\{\matrix{
F_0=\phi_0{n_1-n_0\over n_0+n_1}\hfill\cr
\noalign{\medskip}
G(x)=(\phi_0^2-F_0^2)e^{-(n_0+n_1)\vert x\vert}\hfill\cr
}\right.
\eqno(3.5)
$$
tends to the white noise process
$$\left(\matrix{
<\phi>=F_0\hfill\cr
\noalign{\medskip}
<\phi(x)\phi(0)>=F_0^2+\sigma\delta(x)\hfill\cr
}\right.
$$
in the limit
$$\left(\matrix{
\phi_0\to \infty    \cr
n_0\to \infty    \cr
n_1\to \infty    \cr
}\right.
\qquad\hbox{where}\qquad
\left(\matrix{
\quad F_0=\phi_0{n_1-n_0\over n_0+n_1}\cr
\hbox{\sevenrm and}\hfill\cr
\quad\sigma=2{\phi_0^2\over n_0+n_1}\cr
}\right.
\qquad\hbox{remain constants.}
\eqno(3.6)
$$
It will be also useful to define $\mu={F_0\over\sigma}={n_1-n_0\over 2\phi_0}$.

We will now see that this model is exactly soluble, even though it contains
correlations, because it is a Markovian process.

Note that this two-step process can be easily generalized into a multi-step
process [18].

\noindent {\bf 2 --- Application of the Phase Formalism for $E>0$}

As we have already stressed, $\{\phi(x)\}$ is a Markovian process. Since the
accumulated phase $\theta(x)$ evolves for $E>0$ according to
${d\theta\over dx}=k-\phi(x)\sin 2\theta(x)$, we see that the pair
$\{\phi(x),\ \theta(x)\}$ forms a two-dimensional Markov process.

Let us define
$$\matrix{
\widetilde P_0(\theta,x) \ d\theta\equiv \hbox{the probability to have}
\qquad\left\{\matrix{
\theta(x)\in[\theta,\theta+d\theta]\cr
\hbox{\sevenrm and}\hfill\cr
\phi(x)=+\phi_0\hfill\cr
}\right.\cr
\noalign{\medskip}
\widetilde P_1(\theta,x) \ d\theta\equiv \hbox{the probability to have}
\qquad\left\{\matrix{
\theta(x)\in[\theta,\theta+d\theta]\cr
\hbox{\sevenrm and}\hfill\cr
\phi(x)=+\phi_1\hfill\cr
}\right.
}$$

We can write two coupled master equations for $\widetilde P_0$ and
$\widetilde P_1$
$$\left\{\matrix{
{\partial \widetilde P_0\over \partial x}=-{\partial \over \partial \theta}
[(k-\phi_0\sin 2\theta)
\widetilde P_0]-n_0\widetilde P_0+n_1\widetilde P_1\cr
\noalign{\medskip}
{\partial \widetilde P_1\over \partial x}=-{\partial \over \partial \theta}
[(k-\phi_1\sin 2\theta)
\widetilde P_1]-n_1\widetilde P_1+n_0\widetilde P_0\cr
}\right.
\eqno(3.7)
$$
We now express the probability distributions $P_0(\theta,x)$ and
$P_1(\theta,x)$
of the reduced phase $\theta$, defined modulo $\pi$, in terms of the
probability distributions  $\widetilde P_0(\theta,x)$ and $\widetilde P_1
(\theta,x)$ of the accumulated phase $\theta\in R$ [15]
$$
P_i(\theta,x)=\sum_{n\in Z} \widetilde P_i(\theta+n\pi,x)
$$
$P_0(\theta,x)$ and $P_1(\theta,x)$ are by definition $\pi$-periodic, and
satisfy the same system (3.7) as $\widetilde P_0$ and $\widetilde P_1$. As
$x\to \infty    $, they converge respectively towards stationary solutions
$P_0(\theta)$ and $P_1(\theta)$ of (3.7)
$$\left\{\matrix{
{d\over d\theta}[(k-\phi_0\sin 2\theta)P_0]+n_0P_0(\theta)-n_1P_1(\theta)=0\cr
\noalign{\medskip}
{d\over d\theta}[(k-\phi_0\sin 2\theta)P_1]-n_0P_0(\theta)+n_1P_1(\theta)=0\cr
}\right.
\eqno(3.8)
$$
The sum of these two equations gives a simple relation between $P_0(\theta)$
and $P_1(\theta)$
$$
(k-\phi_0\sin 2\theta)P_0(\theta)+(k-\phi_1\sin
2\theta)P_1(\theta)=\hbox{C}
\eqno(3.9)
$$
where C is a constant which can be evaluated at the point $\theta={\pi\over 2}$
; in fact it is
exactly the density of states $N(E)$ according to (2.18)
$$
 C=kP_0\left({\pi\over 2}\right) + kP_1\left({\pi\over 2}\right)=
kP_{eq}\left({\pi\over 2}\right)=N(E)
\eqno(3.10)
$$
The system (3.8) can now be rewritten as two decoupled equations for
$P_0(\theta)$ and $P_1(\theta)$, containing the a priori unknown constant
$N(E)$
$$\left\{\matrix{
(k-\phi_1\sin 2\theta){d\over d\theta}[(k-\phi_0\sin 2\theta)P_0]
+(n_0+n_1)[k-F_0\sin 2\theta]P_0=n_1N(E)\cr
\noalign{\medskip}
(k-\phi_0\sin 2\theta){d\over d\theta}[(k-\phi_1\sin 2\theta)P_1]
+(n_0+n_1)[k-F_0\sin 2\theta]P_1=n_0N(E)\cr
}\right.
\eqno(3.11)
$$

Finally $N(E)$ will be determined by the normalization conditions that must be
imposed on the $\pi$-periodic solutions $P_0(\theta)$ and $P_1(\theta)$
of (3.11)
$$\left\{\matrix{
\int_0^\pi P_0(\theta)d\theta={n_1\over n_0+n_1}\cr
\noalign{\medskip}
\int_0^\pi P_1(\theta)d\theta={n_0\over n_0+n_1}\cr
}\right.
\eqno(3.12)
$$
where we have used (3.3).

According to (2.19) the Lyapunov exponent can be expressed in terms of the
stationary distributions $P_0(\theta)$ and $P_1(\theta)$
$$
\gamma(E)=<\phi(x)\cos 2\theta(x)>=\int_0^\pi d\theta\cos 2\theta
[\phi_0P_0(\theta)+\phi_1P_1\T]
\eqno(3.13)
$$

The determination of $N(E)$ and $\gamma(E)$ is now reduced to the resolution of
the equations (3.11) for $P_0\T$ and $P_0\T$. The results are presented in
appendix A.
In the following we will only consider various limiting cases for $N(E)$ and
$\gamma(E)$.
\def\arctg{\mathop{\rm arctg}\nolimits}
\def\oppropto{\mathop{\propto}}
\def\fio2{\phi_0^2}

\noindent$\ast$ {\bf Limit $E\to 0^+$}

In order to determine the limiting behaviour of $N(E)$ for $E\to 0^+$, we have
to study how the integrals $I_1$ and $I_2$ given in (A.4) diverge in this
limit. It is convenient to set $\mu_0={n_0\over 2\phi_0},\ \mu_1={n_1\over
2\phi_0}$ and $\mu=\mu_1-\mu_0$.

$\bullet$ For $\mu>0$,\quad $I_1$ has the dominant divergent behaviour and
$N(E)$ vanishes as $E^\mu$
$$
N(E)\opsimeq_{E\to 0}2\sigma{n_1\over n_0}
\left[
{\Gamma(\mu_1)\over\Gamma(\mu)\Gamma(\mu_0)}
\right]^2
\left({E\over\phi_0^2}
\right)^\mu
\eqno(3.14)
$$

$\bullet$ For $\mu<0$,\quad $I_2$ has the dominant divergent behaviour and
$N(E)$ vanishes as $E^{\vert\mu\vert}$
$$
N(E)\opsimeq_{E\to 0}2\sigma{n_0\over n_1}
\left[
{\Gamma(\mu_0)\over\Gamma(\vert\mu\vert)\Gamma(\mu_1)}
\right]^2
\left({E\over\phi_0^2}
\right)^{\vert\mu\vert}
\eqno(3.15)
$$

$\bullet$ For $\mu=0$\quad $I_1$ and $I_2$ diverge only logarithmically and
therefore $N(E)$ vanishes very slowly in comparison to the cases $\mu\neq 0$
$$
N(E)\oppropto_{E\to 0}{1\over(\ln E)^2}
\eqno(3.16)
$$
In the next section (IV) it will be shown how these limiting behaviours of
$N(E)$ can be
understood through a qualitative analysis.

Let us now turn to the Lyapunov exponent $\gamma(E)$. We can study how the
integrals $J_1$ and $J_2$ given in (A.6) diverge in the limit $E\to 0$. $J_1$
(resp. $J_2$) has the dominant divergent behaviour if $\mu>0$ (resp. $\mu<0$)
and we recover the result (2.10)
$$
\gamma(0)=\vert F_0\vert
$$
The case $\mu=0$ deserves special attention since $\gamma(E)$ vanishes as
$$
\gamma(E) \oppropto_{E\to 0}{1\over(-\ln E)}
\eqno(3.17)
$$
This singularity corresponds to the singularity (3.16) found for the density
of states $N(E)$ through the Thouless formula [2]. Note that for $E=0$, the
wavefunctions are not exponentially localized since $\gamma(0)\equiv\lim_{L\to
\infty     }{1\over L}\ln{\rho(L)\over\rho(0)}=0$, but it can nevertheless be
shown that
$\ln{\rho(L)\over\rho(0)}$ behaves asymptotically for large L as $C{L}$ (See
Section IV).

\noindent$\ast$ {\bf Limit of white noise process for $\{\phi(x)\}$}

$\{\phi(x)\}$ tends to a white noise process
$\left(\matrix{
<\phi>=F_0\hfill\cr
<\phi(x)\phi(x')>=F_0^2+\sigma\delta(x-x')\cr
}\right.$
in the limit
$\left(\matrix{
\phi_0\to \infty     \cr
n_0\to \infty     \cr
n_1\to \infty     \cr
}\right.$
provided that
$\left(\matrix{
F_0=\phi_0 \ok {n_1-n_0\over n_1+n_0}\hfill\cr
\sigma=2 \ok {\phi_0^2\over n_0+n_1}\cr
}\right.$
remains constants (3.6).

\noindent The parameter $\mu= \ok {n_1-n_0\over 2\phi_0}={F_0\over\sigma}$ also
remains
constant.
In this limit we recover the results [8,9] for $0<E<\infty     $
$$\left\{\matrix{
N_{WN}(E)= \ok {2\sigma\over\pi^2}{1\over J_\mu^2(z)+N_\mu^2(z)}\hfill\cr
\noalign{\medskip}
{\displaystyle{}^\gamma}{}_{WN}(E)=-\sigma z {d\over dz}\ln
C{J_\mu^2(z)+N_\mu^2(z)}\hfill\cr }\right.\eqno(3.18)
$$
where $z={C{E}\over\sigma}$ and ($J_\mu,\ N_\mu$) are Bessel functions.

\noindent$\ast$ {\bf Limit $E \to \infty$}

In this limit, we recover of course that $N(E)$ behaves asymptotically as the
free density of states (II 1 b))
$$
N(E)\opsim_{E\to \infty }{C{E}\over\pi}
$$
The Lyapunov exponent $\gamma(E)$ vanishes asymptotically as $1\over E$ in
accordance with the result (2.16) for exponentially correlated processes. The
correspondance between (2.15) and (3.5) is the following
$$\left\{\matrix{
\lambda=n_0+n_1\hfill\cr
\lambda\sigma=  {n_0n_1\over(n_0+n_1)^2}\ 8\phi_0^2\hfill\cr
}\right.
$$
and
$$\gamma(E)\opsim_{E\to \infty     }{n_0n_1\over n_0+n_1}
\left({\phi_0^2\over E}\right)
\eqno(3.19)
$$

Let us give a straighforward derivation of this result for the Hamiltonian (See
Fig.2)
$$\eqalign{
H_+&=-{d^2\over dx^2}+\phi^2+\phi'(x)\cr
&=-{d^2\over dx^2}+\phi_0^2+\sum_n(-1)^n2\phi_0\delta(x-x_n)\cr
}$$
where
$\left\{\matrix{
(x_{2n})\quad\hbox{are the points where}\ \phi(x)\ \hbox{changes from}\
(-\phi_0)\ \hbox{to}\ (+\phi_0)\hfill\cr
(x_{2n+1})\quad\hbox{are the points where}\ \phi(x)\ \hbox{changes from}\
(+\phi_0)\ \hbox{to}\ (-\phi_0)\hfill\cr
}\right.$

On the interval [0, L] there will be typically N attractive delta potentials
and
N repulsive delta potentials with (3.3)
$$
L=N(<l>+<h>)=N({1\over n_0}+{1\over n_1})
\eqno(3.20)
$$
The elementary potentials $V_\pm=\phi_0^2\pm 2\phi_0\delta(x)$ have
respectively the following transmission coefficients in the region $E>\phi_0^2$
$$
t_\pm(E)={iC{E-\phi_0^2}\over iC{E-\phi_0^2}\mp\phi_0}
\eqno(3.21)
$$
In the high energy limit, we can neglect interferences between the various
delta potentials and approximate the total transmission coefficient on [0,L]
by a product of elementary coefficients $t\pm(E)$ [16]
$$
\left\vert{\psi(L)\over\psi(0)}\right\vert\opsim_{E\to \infty}
\left\vert t_+(E)t_-(E)\right\vert^{N(L)}
\eqno(3.22)
$$
where
$N(L){\displaystyle\opsim_{L\to \infty }} \ok {L \over { {1\over n_0}+{1\over
n_1}}}$
according to (3.20).
With the help of (3.21) this yields
$$\eqalignno{
\gamma(E)&=-\lim_{L\to \infty     }{1\over L}\ln
\left\vert{\psi(L)\over\psi(0)}\right\vert\cr
&\opsimeq_{E\to \infty     }-{1\over {1\over n_0}+{1\over n_1}}\ln
\left\vert t_+(E)t_-(E)\right\vert&(3.23)\cr
&\opsimeq_{E\to \infty     }{n_0n_1\over n_0+n_1}
\left({\phi_0^2\over E}\right)\cr
}$$

\noindent {\bf 3 --- Application of the Phase Formalism for $E<0$}

{}From the spectral properties of $H_\pm$ we know that $N(E)$ identically
vanishes
 for $E<0$. Nevertheless it is interesting to calculate the Lyapunov exponent
$\gamma(E)$ in this region since it is directly relevant for the problem of
classical diffusion [8]. According to the dynamical equations (2.21), the
Lyapunov
exponent $\gamma(E)$ reads for $E<0$
$$
\gamma(E)=\lim_{L\to \infty     }{1\over L}
\int_0^Ldx[-k\sin 2\t(x)+\phi(x)\cos 2\t(x)]
\eqno(3.24)
$$
and the phase $\t (x)$ evolves according to
$$
{d\t\over dx}=-k\cos 2\t(x)-\phi(x)\sin 2\t(x)
\eqno(3.25)
$$

 Here, there is no need to distinguish
 accumulated and reduced phase, according to the discussion following (2.21).
 As before (3.7), we can easily write Master Equations for $P_0(\t,x)$ and
$P_1(\t,x)$
$$\left\{\matrix{
{\partial P_0\over \partial x}=-{\partial \over \partial
 \t}[(-k\cos 2\t-\phi_0\sin
2\t)P_0]-n_0P_0+n_1P_1\hfill\cr
\noalign{\medskip}
{\partial  P_1\over \partial x}=-{\partial \over \partial
  \t}[(-k\cos 2\t+\phi_0\sin
2\t)P_1]+n_0P_0-n_1P_1\hfill\cr
}\right.\eqno(3.26)
$$
 As $x\to \infty  $, they converge respectively towards stationary solutions
$P_0(\theta)$ and $P_1(\theta)$ of (3.26)
$$\left\{\matrix{
{d \over d    \t}[(k \cos 2\t +\phi_0 \sin
2\t)P_0]-n_0P_0+n_1P_1 =0\hfill\cr
\noalign{\medskip}
{d \over d    \t}[(k\cos 2\t-\phi_0\sin
2\t)P_1]+n_0P_0-n_1P_1 =0\hfill\cr
}\right.\eqno(3.27)
$$
The sum of these two equations gives
$$
(k\cos 2\t+\phi_0\sin
2\t)P_0\T+(k\cos 2\t-\phi_0\sin
2\t)P_1\T=\hbox{constant}
\eqno(3.28)
$$
This constant vanishes at the point $\t={\pi\over 2}$ since
$k[P_0({\pi\over 2})+P_1({\pi\over 2})]=N(E)=0$. The system (3.26) can
now easily be decoupled.

The solutions $P_0(\theta)$ and $P_1(\theta)$, and the calculation
of $\gamma (E)$ in terms of these stationnary distributions
are given in Appendix B.
Finally, we obtain after some transformations
$$
\gamma(E)={n_1\over n_0+n_1}\phi_0C{1+\beta^2}{N(\beta^2)\over D(\beta^2)}
\eqno(3.29)
$$
where
$\left\{\matrix{
\displaystyle{
N(\beta^2)=\int_0^1dx\ x^{\nu_0-1}\left({1-x\over x+\beta^2}\right)^{\nu_1}
{x^2+\beta^2\over(1-x)(x+\beta^2)}}\hfill\cr
\displaystyle{D(\beta^2)=\int_0^1dx\ x^{\nu_0-1}\left({1-x\over x+\beta^2}
\right)^{\nu_1}}\hfill\cr
}\right.
$.

\noindent We have set $\beta={k\over \phi_0}$ , $\nu_1={n_1\over
2\phi_0}{1\over
C{1+\beta^2}}$ and $\nu_0={n_1\over 2\phi_0}{1\over
C{1+\beta^2}}$.

\noindent$\bullet$\quad In the white noise limit (3.6) we recover the result
[8]
$$
{\displaystyle{}^\gamma}{}_{WN}(E=-k^2)=-2\sigma z{K'_{\mu}(z)\over K_{\mu}(z)}
\eqno(3.30)
$$
where $z={k\over\sigma}$ and $K_{\mu}$ is the modified Bessel function.

\noindent$\bullet$\quad In the limit $E\to 0$, we recover (2.10)
$$
\gamma(0)=\vert F_0\vert
$$
In section (VI), we will use the expression (3.29) to obtain the velocity for
the associated problem of classical diffusion in the random force field
$\{\phi(x)\}$.
\def\fio2{\phi_0^2}
\parindent=1.5true cm

\item{\hbox to\parindent{\enskip \bf {IV -}\hfill}}
{\bf LOW ENERGY STATES - A PHYSICAL PICTURE}

The purpose of this section is to provide a simple physical picture that
accounts
 for the low energy part of the spectrum (3.14, 3.16). We will show that the
basic mechanism
is very different from the one which is responsible for the usual Lifchitz
singularities. We have to consider separately the cases $\mu=0$ and $\mu>0$.

\noindent 1) $\mu=0$.

 For a given realization of the disorder, the two
independent solutions of $H_+\psi=0$ have been constructed above (2.11). In
general, none of these solutions is normalizable on the whole line. Instead, if
we consider the problem on a finite interval $[-R,R]$ with Dirichlet boundary
conditions, then one can show that there exists a quasi zero mode (which is a
linear combination of $\psi_0$ and $\psi_1$) whose energy is exponentially
small. One obtains with exponential accuracy [19]
$$
E_0(R)\opsim_{R \to \infty}
\bigg[  {1\over {\int_{-R}^0 {dx\over\psi_0^2(x)}}} +
{1\over {\int_{0}^R {dx\over\psi_0^2(x)}}} \ \bigg]  \ \
{ 1 \over {\int_{-R}^Rdx\ \psi_0^2(x)}}
\eqno(4.1)
$$

Since the corresponding wave function has its support on $[-R,R \ ]$, it is
also a
quasi zero mode on the whole line with same energy. In the presence of
disorder, in order to estimate $E_0$ we will replace $\psi_0(x)$ by its typical
value. When $\{\phi(x)\}$ is a random telegraph process,
the distribution of the
random variable $\int_0^x dy\ \phi(y)$ can be obtained exactly (see Appendix
D).
However, for large $x$, one merely expects that ${1\over\sqrt{\vert
x\vert}}\int_0^x
\phi(y)dy$ will be distributed as a Gaussian process [20,21]. More precisely,
for the case
$\phi_1 = - \phi_0$ and $n_0 = n_1 =n$ the Central Limit
 theorem gives
$\psi_0^{typ} \ok\opsim_{\vert x \vert \to \infty} e^{-\phi_0C{\vert
x\vert\over n}}$.
This gives
$$
E\oppropto_{R\to \infty     }{1\over\sqrt{R}}\ e^{-2\phi_0C{R\over n}}
\eqno(4.2)
$$
Phrased differently, this means that a quasi zero mode of energy $E$ has
typically a
spatial extension $2R$ such that
$$
\ln E \sim -2\phi_0C{R\over n}
\eqno(4.3)
$$
Therefore the number of such states per unit length behaves typically as
$$
N(E) \sim {1\over 2R}\sim{1\over\ln^2E}
\eqno(4.4)
$$
This argument can be easily generalized to account for other types of
correlations. For instance, if one assumes that the correlation function has a
power law behaviour
$$\eqalign{
<(\phi(x)-\phi(y))^2&>\sim\vert x-y\vert^\alpha\qquad(\alpha>0)\cr
&\vert x-y\vert\to \infty  \cr
}\eqno(4.5)
$$
 one can easily show along the same lines that
$$
N(E)\oppropto_{E \to 0} {1\over {(-\ln E)^{2\over 2+\alpha}}}
\eqno(4.6)
$$
This last result is in agreement with the one obtained in [22] in the context
of anomalous diffusion.

\noindent 2) $\mu>0$.

 For simplicity we consider the case where $\phi_1=-\phi_0$
with $n_1\gg n_0$. The profile of the potential is sketched on Fig. 2. Each
time
$\phi(x)$ jumps from $\phi_0$ to $-\phi_0$ and then back from $-\phi_0$ to
$\phi_0$, there appears a potential well $V_-(x)=-2\phi_0\delta(x)$ which is
then
followed by a potential barrier  $V_+(x)=2\phi_0\delta(x)$. Such a dipole like
configuration can support localized states which account for the low energy
behaviour of the density of states. Since $n_1\gg n_0$ the average distance
between such
configurations is large. Therefore one can safely ignore interference effects
and consider a single doublet of width a. The potential is given by
$$
V(x)=\fio2-2\phi_0\delta(x)+2\phi_0\delta(x-a)
\eqno(4.7)
$$
An elementary calculation shows that this potential can support a bound state
provided
$$
{E\over\fio2}=e^{-2a\phi_0C{1-{E\over\fio2}}}
\eqno(4.8)
$$
Here we deal with low energy states such that $0<E\ll \fio2$, this gives
$$
E\simeq\fio2 e^{-2a\phi_0}
\eqno(4.9)
$$
which is consistent with the previous inequality if $a\phi_0\gg 1$.
If we would just consider the attractive potential well
$$
V(x)=\fio2-2\phi_0\delta(x)
$$
this would lead to a zero energy state $\psi_0(x)\propto e^{-\vert
x\vert\phi_0}$ (consistent with supersymmetry). It is the coupling of this
state
with the repulsive potential barrier  $2\phi_0\delta(x-a)$ that increases the
energy up to $E=\fio2 e^{-2a\phi_0}$. This picture is indeed confirmed by an
elementary calculation to first order in perturbation theory. Since the width
of
the doublet is distributed according to the distribution
$$p(a)= \theta (a) \ n_1 \ e^{-n_1a}
\eqno(4.10)
$$
this implies that the number of such states per unit length is
$$
N(E)\oppropto_{E\to 0} e^{ \ok -{n_1\over 2\phi_0}\ln{\fio2\over E}}=
\left(
{E\over \fio2}
\right)^{n_1\over 2\phi_0}
\eqno(4.11)
$$
Since $\mu={n_1-n_0\over 2\phi_0}\simeq{n_1\over 2\phi_0}$ this result is
consistent with eq. (3.14).

The above discussion clearly illustrates that the basic mechanism which is
responsible for the low energy behaviour is the existence of low energy states
that would have strictly zero energy if one could ignore the couplings between
wells. This mechanism is very different from the one which is at work in the
case
of Lifchitz singularities. In this case, very small energy can occur only if
large regions of space are free from impurities. This is just the opposite
picture of what happens in our case.

\noindent This discussion can be generalized to the case $\phi_1<0<\phi_0$ with
$<\phi>\neq 0$. Assuming $0<\vert\phi_1\vert<\phi_0$ and $n_0=n_1=n$, an exact
calculation of the density of states gives again a power law behaviour
$$
N(E)\oppropto_{E\to 0} E^{ \ok {n\over 2}\left({1\over\vert\phi_1\vert}-
{1\over\vert\phi_0\vert}\right)}\ .
\eqno(4.12)
$$

However the situation is completely different when $\phi_1=0$. In this case, a
single doublet configuration of length $a\gg{1\over\phi_0}$
$$
V(x)=\fio2[\t(x-a)+\t(-x)]-\phi_0\delta(x)+\phi_0\delta(x-a)
\eqno(4.13)
$$
supports a low energy state
$$
E(a)\simeq\left({\pi\over 2a}\right)^2\ .
\eqno(4.14)
$$
The dependence on the length a is very different from (4.9), and (4.10) gives
now
a Lifchitz singularity
$$
N(E)\oppropto_{E\to O^+}\exp \bigg(-{\pi\over 2}{n_1\over\sqrt{E}} \bigg)
\eqno(4.15)
$$
This result is in agreement with an exact calculation of the density of states
that we do not
reproduce here.

\noindent For $0<\phi_1<\phi_0$, one can prove that $H_+$ is bounded from below
by $\phi_1^2$; $N(E)$ starts now with a Lifchitz singularity at $\phi_1^2$
$$
\eqalign{
N(E)&=0\quad E<\phi_1^2\cr
N(E)&\oppropto_{E\to(\phi_1^2)^+}\exp \bigg(-{\pi\over 2}
{n_1\over\sqrt{E-\phi_1^2}}\bigg) \ .
}\eqno(4.16)
$$
\def\oppropto{\mathop{\propto}}
\def\arctg{\mathop{\rm arctg}\nolimits}
\def\fio2{\phi_0^2}
\parindent=1.1true cm
\item{\hbox to\parindent{\enskip \bf {V -}\hfill}}
{\bf COMPUTER SIMULATIONS AND NUMERICAL CALCULATIONS}

In this section, we consider various probability distributions of the lengths
steps and discuss the density of states $N(E)$ and inverse localization length
$\gamma(E)$ as obtained from simulations and/or numerical calculations.

The procedure to obtain the integrated density of states $N(E)$ by simulation
for
the model (III.1) is the following. We generate the random process $\{\phi\}$
with
the laws (3.1) for the lengths of intervals where $\phi(x)$ is constant. Then
for
each energy E, we integrate the dynamical equation (2.5) for the total
accumulated phase $\t_E(x)$
$$
{d\t_E\over dx}=k-\phi(x)\sin 2\t_E(x)
$$
and we finally compute $N(E)$ with (2.2)
$$
N(E)=\lim_{L\to \infty }
\left({\t_E(L)-\t_E(0)\over \pi L}\right)\ .
$$
$N(E)$ is a self averaging quantity and requires therefore only one
configuration
$\{\phi\}$ to be computed. We have generated typically $3.10^5$ intervals, and
we
have checked with a great accuracy that the $N(E)$ obtained was indeed
independent of the particular realization of the disorder $\{\phi\}$.

In order to have a better accuracy, we did not integrate the dynamical equation
for $\t(x)$ step by step, but we rather integrated it exactly on each interval
where $\phi(x)$ is constant. For an interval of length $\ell$ on which
$\phi(x)=+\phi_0$, the initial value $\t_i$ and the final value $\t_f$ of the
phase $\t(x)$ are related through
$$
\ell=\int_{\t_i}^{\t_f}{d\t\over k-\phi_0\sin 2\t}
\eqno(5.1)
$$
The integration requires to separate $k<\phi_0$ from $k>\phi_0$ (See appendix
C).

In Fig. 3, we display $N(E)$ for $\phi_0=-\phi_1=1$ and various values of $n =
n_0 =n_1$.
 Notice that, as expected, $N(E)$ is a continuous function of $E$. However, it
seems
that, in general, its derivative is not.

\noindent For large $n$ (Fig.3a), $N(E)$ is close to the free case
${C{E}\over\pi}$
except for very small $E$ (3.16). This is easily understood if we consider
two successive steps of small lengths $\left(\approx{1\over n}\right)$ in
(2.5).
After two steps the phase increase is
$$
d\t_0 + d\t_1 \approx  ( k-\phi_0\sin 2\t ) dx_0  +( k+\phi_0\sin 2\t )
dx_1\approx \ kdx\,
\eqno(5.2)
$$
This leads to a quasi free density of states.

\noindent In contrast, small values of $n$ (Fig.3d) lead to \hfill\break
$$\matrix{
i)\ \ \hbox{a plateau}\quad N(E)\approx {n\over 2}\quad \hbox{if}
\quad E<\fio2\hfill\cr
ii)\ \hbox{a parabolic behaviour}\quad N(E)\approx{C{E-\fio2}\over\pi}
\quad \hbox{if}\quad E>\fio2\hfill\cr
}\eqno(5.3)
$$
The behaviour $i)$ can be explained by noting that, for small $n$,
 $\t_E(L)-\t_E(0) \sim N{\pi\over 2}$.
Therefore after N steps, we get
$$
N(E)\approx \lim_{N,L\to \infty     } {1\over\pi L}\left({\pi\over 2}
N\right)={n\over 2}
\eqno(5.4)
$$
For $E>\fio2$, one can show (see Appendix C) that $\t(L)\approx{\pi
L\over\tau}$, hence we get $ii)$
$$
N(E)\approx {1\over\tau}={C{E-\fio2}\over\pi}
$$
A quasi-free behaviour develops only for $E>\fio2$.

Now we turn to the Lyapunov exponents and consider the ending angles
$\pi$-periodic stationary distributions. $W_0\T \  d\t$ (resp. $W_1\T  \ d\t$)
is the probability for
the reduced phase to be in the interval $[\t,\t+d\t]$ when the system switches
from state $\phi_0$
to state $\phi_1$ (resp. from $\phi_1$ to $\phi_0$). For $E>0$, (2.3) and (2.5)
lead to
$$
\gamma(E)=-{n_0n_1\over 2(n_0+n_1)}\ .\
\int_{-{\pi\over 2}}^{+{\pi\over 2}}
 \ln \left\vert{C{E}-\phi_0\sin 2\t\over\sqrt{E}-\phi_1\sin 2\t}
\right\vert\ .\
[W_0\T-W_1\T] d\t
\eqno(5.5)
$$
The stationarity condition implies the relationship
$$
W_0\T=\int_0^\infty     d\ell\int_{-{\pi\over 2}}^{+{\pi\over
2}}d\t'f_0(\ell)W_1(\t')\delta
(\t-g(\ell,\t'))
\eqno(5.6)
$$
(g deduced from (5.1)) and, after some algebra, the differential equations
$$\eqalign{
{d\over d\t}
\big[ W_0\T(k-\phi_0\sin 2\t) \big]=n_0(W_1\T-W_0\T)\cr
{d\over d\t}
[W_1\T(k-\phi_1\sin 2\t)]=n_1(W_0\T-W_1\T)\cr
}\eqno(5.7)
$$
Comparison with (3.8) shows that distributions $W_{0,1}\T$ and $P_{0,1}\T$ are
the same up to a normalization constant
$$
P_{0,1}={n_{1,0}\over n_0+n_1}\ W_{0,1}
\eqno(5.8)
$$
In particular, we can demonstrate, via (5.8), the equivalence of the two
calculations, (3.13) and (5.5), of the Lyapunov exponent $\gamma(E)$.
(5.8) is somewhat unexpected and could be possibly related to the Poisson
process we use for the step lengths.

Distributions $W_0\T$ and $W_1\T$ are plotted in Figs. 4 and 5 for two values
of $n$.
$W_0\T$ (resp. $W_1\T$)  exhibits a peak at
$\t=\alpha_0$ (resp. at $\t=\alpha_1$) , the magnitude of which is a decreasing
function of $n$. For
small $n$ (Fig.4), $W_0$ is essentially concentrated around $\alpha_0$.
 On the other hand, for
large $n$ (Fig.5), $W_0$ becomes practically flat. (5.5) shows that, in this
case, we
must get $\gamma(E)\approx  0$ (free case). The small bump at $\t=\alpha_1$
(= $\alpha_0-{\pi\over 2}$ when $\phi_0=-\phi_1$) is essentially a memory
effect
due to the preceding step ($\phi(x)=\phi_1$).
With the aid of (3.13) and (5.5, 8), we could compute $\gamma(E)$
numerically as well as by simulations. Some results are displayed in Fig. 6
($n_0=n_1$)
and Fig. 7 ($n_0=2n_1$). We
clearly see that $\gamma(0)=\vert F_0\vert$ (2.10) and that $\gamma(E)$, with
$E$
fixed, is a decreasing function of $n_0$ (at least for $E$ not too large). This
is in agreement with the above discussion. The predicted asymptotic behaviour
$\gamma(E) \ok \oppropto_{E\to \infty }E^{-1}$ (3.19) is reached for values
that largely
depend on $n_0$ and $n_1$. For instance, with $n_0=1=2n_1$, it is practically
attained for  $E\gsim 2$. But when $n_0= 2 n_1 =4$, values of $E\gsim 20.-30.$
are needed.

Small $n_{0,1}$ values deserve a special comment. As already noticed,
$W_0(W_1)$
distribution is peaked at $\alpha_0(\alpha_1)$. Approximating $P_{0,1}\T$ by
$\delta$-functions, (3.13) readily leads to ($\phi_0=-\phi_1)$
$$\matrix{
&\gamma(E)\approx C{\fio2-E}\quad&,\quad E<\fio2\cr
\noalign{\medskip}
&\gamma(E)\approx 0\hfill&,\quad E\geq\fio2\cr
}\eqno(5.9)
$$
This is precisely what we get by means of computer simulations for
$n_0=0.01=2n_1$. However, eq.(2.10) $\gamma(0)=\vert F_0\vert$ seems
contradictory
with (5.9). Investigating the limit $E\to 0^+$ numerically, we could see that
very small $E$ values are needed ($\approx 10^{-3}-10^{-4}$) to detect a
decrease
of $\gamma(E)$ towards its limit $\vert F_0\vert$. So, the behaviour of the
Lyapunov exponent at small energy appears to be far from trivial.

We now consider step length distributions whose first moment diverges. More
precisely, exponential laws $f_{0,1}(\ell)$ of eq.(3.1) are replaced by
distributions
who behave as
$$
\ell^{-(1+\alpha)}\quad\hbox{for}\quad \ell\to \infty     \quad(0<\alpha\le 1)\
{}.
\eqno(5.10)
$$
$\alpha=1$ corresponds to a one-sided Cauchy law.

\noindent Detailed analytical calculations such as those performed in the
preceeding
sections become hardly feasible. However, some simple characteristic features
appear in the computer simulations. For instance, when
$\phi_0=-\phi_1$, the density of states, $N(E)$, is given by
$$
N(E)\approx\t(E-\fio2) {{C{E-\fio2}} \over \pi}
\eqno(5.11)
$$

\noindent In fact, one can prove that result (5.11) is an exact one. Indeed,
since
the first moment of (5.10) is infinite, the sum L, of N independent positive
random variables all distributed according to (5.10) behaves, when $N\to \infty
$, like
[23]
$$\matrix{
&L\sim N^{1/\alpha}\quad&,\quad \alpha<1\cr
\noalign{\medskip}
&L\sim N\ln N\quad&,\quad \alpha=1\cr
}\eqno(5.12)
$$
More precisely, ${L\over N^{1/\alpha}}$ (or ${L\over N\ln N}$ if $\alpha=1$) is
distributed
according to a L\'evy stable law.

\noindent Thus, for $E<\fio2$, we get
$$\matrix{
N(E)\psim {1\over\pi}\ .\ {{{\pi\over 2} N} \over {N^{1/\alpha}}} \ok
\operarrow_{N\to \infty } 0
\quad (\alpha<1)\cr
\noalign{\medskip}
N(E)\psim {1\over\pi}\ .\ {{{\pi\over 2} N} \over {N\ln N}} \ok
\operarrow_{N\to \infty } 0
\quad (\alpha=1)\cr
}\eqno(5.13)
$$
Now we examine what happens for $E>\fio2$ and call
$\ell_1,\ell_2,\cdots,\ell_N$
the lengths of the successive steps
$$
L=\ell_1+\cdots+\ell_N\sim N^{1/\alpha},\quad(\alpha<1,\ N\to \infty )
\eqno(5.14)
$$
According to (C.4), each $\ell_i$ can be written as
$$\eqalign{
&\ell_i=n_i\tau+\delta\ell_i\ ,\cr
&0<\delta\ell_i<\tau\cr
}\eqno(5.15)
$$
and therefore
$$
L=\tau(\sum_in_i)+\sum_i\delta\ell_i
\eqno(5.16)
$$
The upper bound $\displaystyle\sum_i\delta\ell_i<N\tau$ allows to neglect
$(\displaystyle\sum_i
\delta\ell_i)$ in L, and we finally get
$$\matrix{
L=(n_1+\cdots+n_N)\tau+O(N^{1/\alpha})\hfill\cr
\ecart
\t(L)=(n_1+\cdots+n_N)\pi+O(N^{1/\alpha})\hfill\cr
\ecart
N(E)={1\over\tau}={C{E-\fio2}\over\pi}\quad E>\fio2\ .\hfill\cr
}\eqno(5.17)
$$
Replacing $N^{1/\alpha}$ by $N\ln N$ in (5.14-17), we easily show that the
final
result (5.17) is still valid when $\alpha=1$. Moreover, Thouless's formula [2]
implies that
$$
\gamma(E)=0\quad\hbox{when}\quad E>\fio2
\eqno(5.18)
$$
Thus, the case $\alpha\leq 1,\ \phi_0=-\phi_1$, is like a "free" one, up to the
shift
of $\fio2$ in energy.

However, simulations with $\phi_0\not=-\phi_1$ reserve some surprise. In Fig.
8,
three different events drawn with the same law, $\alpha=0.5$ are displayed.
We now notice that $N(E)$ is no longer a self-averaging quantity. In (5.14)
splitting $L$ into
$L_0$ (and $L_1$), the sum of all the steps lengths where $\phi(x)=\phi_0$ (or
$\phi_1$) and observing that $L_0$ and $L_1$ are both of the same order
($N^{1/\alpha}$ or $N\ln N$), we get, in analogy with (5.14-17)
$$
N(E)={L_0\t(E-\fio2)C{E-\fio2}+L_1\t(E-\phi_1^2)C{E-\phi_1^2}\over\pi(L_0+L_1)}\ ,
\eqno(5.19)
$$
in perfect agreement with our simulations. Of course, $L_0$ and $L_1$ now
strongly
depend on each drawing.
\noindent It is interesting to point out that \hfill\break
i) $N(E)\sim{C{E}\over\pi}\quad\hbox{when}\quad E\to \infty     $\hfill\break
ii) $\fio2=\phi_1^2$ in (5.19) leads to
$N(E)={\t(E-\fio2)C{E-\fio2}\over\pi}$, i.e. $N(E)$ is again self-averaging.

To conclude this section, we mention that nothing special seems to happen when
$1<\alpha<2$. In  particular, (5.11, 19) are not observed and the general
behaviour of $N(E)$ is quite similar to the one we obtained with the
exponential
law.
\def\oppropto{\mathop{\propto}}
\def\arctg{\mathop{\rm arctg}\nolimits}
\def\fio2{\phi_0^2}

\parindent= 1.5 true cm
\item{\hbox to\parindent{\enskip \bf {VI -} \hfill}}
{\bf SOME PHYSICAL APPLICATIONS}
\parindent= 1 true cm

In this section we describe some physical problems to which the above model can
be applied. As mentionned in the introduction, one of the most interesting
application is the description of classical diffusion in a one dimensional
random medium. In the presence of a force $\phi(x)$, the probability density
$P(x\ t\ \vert\ y\ 0)$ satisfies the Fokker-Planck equation
$$
{\partial  P\over \partial t}={\partial \over \partial x}
\left({\partial P\over \partial x}-2\phi(x)P\right)
\eqno(6.1)
$$
supplemented by the initial condition $\displaystyle P(x\ t\ \vert\ y\ 0)
\operarrow_{t\to 0^+} \delta(x-y)$

\noindent As we have already explained in the introduction, the transformation
$$
P(x\ t\ \vert\ y\ 0)=e^{\int_y^x\phi(z)\ dz} G_S(x\ t\ \vert\ y\ 0)
\eqno(6.2)
$$
allows to rewrite (6.1) as the Schr\"odinger problem
$$\left\{\eqalign{
&{\partial G_S \over \partial t}=-H_+ G_S\cr
& G_S (x\ t\ \vert\ y\ 0)\operarrow_{t\to 0^+} \delta(x-y)\cr
}\right.\eqno(6.3)
$$
where $H_+=-{d^2\over dx^2}+\phi^2(x)+\phi'(x)$ is the Hamiltonian whose
properties have been discussed in the preceding sections.

\noindent The Laplace transform
$\displaystyle\widehat G_S (x,y;E)=\int_0^\infty  dt\ e^{-Et} \  G_S
(x\ t\ \vert\ y\ 0)$ satisfies
$$
(E+H_+) \ \widehat G_S (x,y;E)=\delta(x-y)
\eqno(6.4)
$$
The Laplace transform
$\ok \widehat P(x,y;E)=\int_0^\infty  dt\ e^{-Et}P(x\ t\ \vert\ y\ 0)$
therefore reads
$$
\widehat P(x,y;E)=e^{\int_y^x\phi(z)\ dz}<x\ \vert\ {1\over H_++E}\ \vert\ y>
\eqno(6.5)
$$
When increasing the mean bias $<\phi>$, it is known that one obtains a
succession of phases [10] characterized by different anomalous diffusive
behaviour.
 The characterization of these phases requires the knowledge of the
velocity and the diffusion constant. Here we just consider the velocity given
by [8]
$$
{1 \over V}=\lim_{E\to 0^+}< \widehat P(x,x;E)>
\eqno(6.6)
$$
In order to express $V$ in terms of the localization problem, we consider the
average of the resolvant at coinciding points
$$
\lim_{L\to \infty }{1\over L}\int_0^Ldx<x\ \vert\ {1\over H_+-{\cal E} -i
\epsilon}
\vert\ x> =-{d\over d\cal E}\ \gamma({\cal E})+i \pi\rho(\cal E)
\eqno(6.7)
$$
For negative $\cal E$, the density of states $\rho(\cal E)$ vanishes and (6.7)
reduces for $E>0$ to
$$
\lim_{L\to \infty       }{1\over L}\int_0^Ldx<x\ \vert\ {1\over H_++E}\ \vert\
x>
={d\over dE}\ \gamma(-E)
\eqno(6.8)
$$
The velocity is therefore given in terms of the left derivative of the Lyapunov
exponent at zero energy
$$
{1\over V}=\lim_{E\to 0^+}< \widehat P(x,x;E)>=\lim_{E\to 0^+}{d\over dE}\
\gamma(-E)
\eqno(6.9)
$$

For the model of rectangular barriers of random lengths described in (III.1) we
have obtained the Lyapunov exponent in the negative energy region in
(3.29).\hfill\break
It follows that the velocity displays a transition at $\mu=1$
$$\left\{\eqalign{
&V=0\quad\hbox{if}\quad\vert\mu\vert<1\cr
&V={2\sigma(\vert\mu\vert-1)\over1-{\sigma^2\over\fio2}\ \vert\mu\vert}
\quad\hbox{if}\quad\vert\mu\vert>1\cr
}\right.\eqno(6.10)
$$
In the white noise limit (Eq. 3.6), we recover the result [8]
$$\left\{\eqalign{
&V=0\quad\hbox{if}\quad\vert\mu\vert<1\cr
&V=2\sigma(\vert\mu\vert-1)
\quad\hbox{if}\quad\vert\mu\vert>1\cr
}\right.\eqno(6.11)
$$

It is interesting to rederive this result by performing a direct
configurational average in (6.6). The zero energy Green function at coinciding
points $P(x,x;0^+)$ can be written for any configuration of the force field
$\{\phi(x)\}$
$$
P(x,x;0^+)=2\int_x^{+\infty }d\xi\ e^{-2\int_x^\xi\phi(u)du}
\eqno(6.12)
$$
provided one assumes $F_0>0$ [8].
Averaging over the disorder $\{\phi(x)\}$ gives
$$
{1\over V}=2\int_0^{+\infty }dL< e^{-2\int_0^L \phi(u)du}>
\eqno(6.13)
$$

For the two-step model (III.1), $\phi(x)$ can only take two values
($\pm\phi_0$), therefore
$$
\int_0^L\phi(u)du=\phi_0\ A(L)-\phi_0[L-A(L)]
\eqno(6.14)
$$
where $A(L)$ is the random variable that measures the total length in the
state\break $\phi(x)=+\phi_0$ during the interval $[0,L]$.\hfill\break
The velocity V can thus be expressed in terms of the probability density
${\cal P}_L(x)dx=Pr(x<A(L)<x+dx)$. One obtains
$$\eqalignno{
{1\over V}&= 2\int_0^{\infty} dL<e^{2\phi_0L-4\phi_0A(L)}>\cr
&=2\int_0^{\infty} dL\ e^{2\phi_0L}\int_0^Ldx\ e^{-4\phi_0x}{\cal P}_L(x)\ .
&(6.15)\cr
}$$
{}From the expression of ${\cal P}_L(x)$ given in Appendix D one recover
 (6.10) after some tedious calculations.

Another interesting application of this model is the fact that it can be used
to discuss the spectral properties of the two dimensional Euclidean Dirac
operator in a special type of random magnetic field .
Consider the two dimensional Euclidean Dirac operator
$$
iD\mkern -10mu /=i\sigma_1
\left({\partial \over \partial t}+iA_t \right) +i\sigma_2
\left({\partial \over \partial x}+iA_x \right)
\eqno(6.16)
$$
coupled to the external field
$$\left\{\eqalign{
&A_t=f(x)\cr
&A_x=0\cr
}\right.\eqno(6.17)
$$
If we set $\psi(x,t)=\exp i{\omega} t\ \chi(x)$ where $\chi(x)=(v(x), u(x))$ is
a two
component spinor, the eigenvalue equation
$$
iD\mkern -10mu / \psi=k\psi\ .
\eqno(6.18)
$$
then reads
$$\left\{\eqalign{
&{du\over dx}-u\phi=kv\cr
-&{dv\over dx}-v\phi=ku\cr
}\right.\eqno(6.19)
$$
where
$$
\phi(x)=f(x)+\omega
\eqno(6.20)
$$

After decoupling one obtains
$$\pmatrix{
H_-&0\cr0&H_+\cr}
\ \chi=k^2\chi
\eqno(6.21)
$$
The knowledge of the density of states of the one dimensional model therefore
permits to evaluate the density of states in the quasi one dimensional random
magnetic field
$$
B={\partial A_t \over {\partial x}}  - {\partial A_x \over {\partial t}} =f'(x)
\eqno(6.22)
$$
In a previous paper [24] we have shown that the low energy part of $\rho(E)$
is enhanced with respect to the free case. Although the analytical expressions
are of course model dependent, the enhancement effect is expected to occur on
very general grounds.

The type of disorder that we have introduced in section III can be motivated by
the following picture. Imagine that the potential $A_t(x)$ is created by a line
of dipoles of same strength distributed randomly on the $x$ axis with
alternating signs $+2f_0, -2f_0, +2f_0, -2f_0\cdots$ The resulting potential
$\ok f(x)=-f_0\sum_i\epsilon(x-y_i)$
is an ensemble of square functions such that each sample function can only take
the value $\pm f_0$. If one assumes further that the probability to find $m$
dipoles on an interval of length $x$ is given by the Poisson process
$$
p(m,x)={(nx)^m \over {m!}} \exp(-nx)
\eqno(6.23)
$$
then $f(x)$ is a random telegraph process. After some calculations, very
similar
to the one presented in [24] one obtains
$$
N(E)={1\over \pi}\int_{-\infty }^{+\infty }N(\phi_0=f_0+\omega \ ,
\phi_1=-f_0+\omega \ ;E)\ d\omega
\oppropto_{E\to 0^+}{1\over \left(-\ln{E\over f_0^2}\right)^3}
\eqno(6.24)
$$
Similar logarithmic behaviours have been obtained by a variational method in
the context of the Schwinger model [25]. However a direct comparison with our
result
 is not possible because the gauge field measure is not the same.

Interestingly enough, this two dimensional Dirac operator can also model other
systems. For instance it arises in the study of the electronic properties of
polyacetylene. In this context $\phi(x)$ is proportional to the dimerization
pattern of the carbon-hydrogen chain [26,27]. Introduction of the disorder can
of
course influence the nature of the ground state [28]. It can also model
certain layered structures with a piecewise constant order parameter.
Application of these ideas to Fermi superfluids was recently discussed [14].

Another potential application of this model is in the field of dynamical
systems perturbed by noise. A large amount of works have been devoted to the
study of two dimensional linear stochastic systems of the form
$$
\dot x(t)=Ax(t)+\eta(t)Bx(t)
\eqno(6.25)
$$
where A and B are 2x2 matrices , $\eta(t)$  a random noise and
$x(t)=\pmatrix{x_1\cr x_2\cr}$
 is a two dimensional vector. In order to understand how the linear system
$\dot x(t)=Ax(t)$ is perturbed by the noise in the asymptotic regime $t\to
\infty $
one introduces the Lyapunov exponent
$$
\lambda=\lim_{t\to \infty  }{1\over t}\log \vert\vert x(t)\vert\vert
\eqno(6.26)
$$
which characterizes the stability of the solution and the rotation number
$$
\rho=\lim_{t\to \infty }{1\over t}\arctg {x_2(t)\over x_1(t)}
\eqno(6.27)
$$
Although there are general theorems regarding the existence of such limits, in
most cases one cannot compute them analytically. Therefore one has to resort to
perturbative methods (for instance small noise expansions) [29].

The Dirac equation (6.19) can be written in this canonical form with the
identification
$$
A=C{E}\pmatrix{0&1\cr-1&0\cr}\qquad
B=\pmatrix{1&0\cr0&-1\cr}
\eqno(6.28)
$$

A more physical picture is to consider that
$x(t)=\pmatrix{p\cr q\cr}$
characterizes the two dimensional phase space of an harmonic oscillator
perturbated by a random telegraph process. This linear system (6.25) can be put
 into an hamiltonian form with
$$
H={C{E}\over 2}(p^2+q^2)-\phi(t)pq
\eqno(6.28)
$$
{}From the discussion in section III it follows that $\lambda$ and $\rho$ are
nothing but the inverse localization length and density of states of the
localization problem.
\parindent= 1.5 true cm
\item{\hbox to\parindent{\enskip \bf {VII -} \hfill}}
{\bf CONCLUSION}
\parindent=1 true cm

In this work we have studied a model of localisation based on the Witten
Hamiltonian
of supersymmetric quantum mechanics. This model exhibits interesting behaviour
of the
density of states  at low energy. We have shown how this behaviour can be
understood through
simple physical arguments. In the case $\mu=0 $, it is the existence of
quasi-zero modes that is
responsible for the logarithmic singularity, whereas for $\mu \ne 0$ it is the
formation of low
energy  bound states that explains the power law behaviour. These mechanisms
rely of course
 crucially on the supersymmetric structure of the potential. The resulting
expressions are
 qualitatively similar to the ones that had been obtained before in the case of
a white noise.
 In that respect, the existence
of a finite correlation length does not make much difference. In contrast, when
the lengths of
the barriers are drawn with a distribution with infinite moments,
we have demonstrated, analytically as well as numerically,
that the density of states develops a very different behaviour. A similar
observation
was  recently made in the context of anomalous diffusion. It was shown in [30]
that the asymptotic
 regime can be qualitatively different in the presence of correlated transfert
rates (see also [12]).
 An extension
of this model in higher dimension is certainly worth pursuing. In particular it
would be interesting
to extend the zero mode analysis to the case of the random magnetic field
problem.
\vskip 1.5 cm
\def\arctg{\mathop{\rm arctg}\nolimits}
\def\oppropto{\mathop{\propto}}
\def\fio2{\phi_0^2}

\noindent {\bf APPENDIX A} : CALCULATION OF N(E) AND $\gamma$(E) FOR $E>0$

The determination of $N(E)$ and $\gamma(E)$ is reduced to the resolution of
the equations (3.11) for $P_0\T$ and $P_0\T$, which contain the expressions
$(k-\phi_0\sin 2\t)$ and $(k-\phi_1\sin 2\t)$.
The solutions will be different if these expressions can vanish or not for
some values of $\t$, and it is therefore necessary to separate the cases
$E>\phi_0^2$ from $E<\phi_0^2$ and $E<\phi_1^2$ from $E>\phi_1^2$.

In order to simplify the discussion, we will set $\phi_1=-\phi_0$, in which
case the spectrum is only divided into two regions $0<E<\phi_0^2$ and
 $E>\phi_0^2$.
\medskip
\boxit{3pt}{a) Results for $0<E<\phi_0^2$
$\left(\hbox{in the case}
\left\{\matrix{
\phi_0>0\hfill\cr
\phi_1=-\phi_0\hfill\cr
}\right.\right)
$}

We introduce the angle $\alpha\in ]0,{\pi\over 4}[$ such that
$\sin 2\alpha=C{E\over \phi_0^2}$ and the function\break
$q\T=\left\vert{\cos(\t-\alpha)\over \sin(\t+\alpha)}\right\vert^{\nu_1}
\left\vert{\sin(\t-\alpha)\over \cos(\t+\alpha)}\right\vert^{\nu_0}$
where
$\nu_1={n_1\over 2\phi_0\cos 2\alpha}$ and
$\nu_0={n_0\over 2\phi_0\cos 2\alpha}$.

In this region of the spectrum the resolution of (3.11) requires some care
since the expressions $(k\pm\phi_0\sin 2\t)$ can vanish and this is why the
$\pi$-periodic solutions $P_0\T$ and $P_1\T$ are defined by two different
expressions on the intervals $[\alpha,{\pi\over 2}+\alpha]$ and
$[{\pi\over 2}+\alpha, \pi+\alpha]$
$$\left\{\matrix{
{\scriptstyle \alpha\leq \t\leq {\pi\over 2}+\alpha}\hfill&P_0\T={n_1\over
4\phi_0^2}N(E)
{q\T\over \cos(\alpha+\t)\sin(\alpha-\t)}
\int_{{\pi\over 2}-\alpha}^\t
{dt\over\cos(\alpha-t)\sin(\alpha+t)q(t)}\cr
\noalign{\bigskip}&&(A.1)\cr
{\scriptstyle{\pi\over 2}+\alpha\leq\t\leq\pi+\alpha}\hfill&P_0\T={n_1\over
4\phi_0^2}N(E)
{q\T\over \cos(\alpha+\t)\sin(\alpha-\t)}
\int_{\pi-\alpha}^\t
{dt\over\cos(\alpha-t)\sin(\alpha+t)q(t)}\cr
}\right.
$$

$$\left\{\matrix{
{\scriptstyle \alpha\leq \t\leq {\pi\over 2}+\alpha}\hfill&P_1\T={n_0\over
4\phi_0^2}N(E) {q\T\over \cos(\t-\alpha)\sin(\t+\alpha)}
\int_{{\pi\over 2}-\alpha}^\t
{dt\over\cos(\alpha+t)\sin(\alpha-t)q(t)}\cr
\noalign{\bigskip}&&(A.2)\cr
{\scriptstyle{\pi\over 2}+\alpha\leq\t\leq\pi+\alpha}\hfill&P_1\T={n_0\over
4\phi_0^2}N(E) {q\T\over \cos(\t-\alpha)\sin(\t+\alpha)}
\int_{\pi-\alpha}^\t
{dt\over\cos(\alpha+t)\sin(\alpha-t)q(t)}\cr
}\right.
$$

The functions $P_0\T$ and $P_1\T$ are presented on Figs. 4 and 5 for a certain
choice of the parameters $\phi_0,\ n_0,\ n_1$.

The normalization condition
$\int_\alpha^{\alpha+\pi}P_0\T d\t={n_1\over n_0+n_1}$ gives the density of
states $N(E)$
$$
N(E)={2\sigma\over I_1+I_2}
\eqno(A.3)
$$
where $I_1$ and $I_2$ are double integrals that read after some transformations
$$\eqalignno{
I_1&=(1+\beta^2)\int_0^\infty    dx\ {x^{\nu_0-1}\over(x+\beta^2)^{\nu_1}}
{1\over\vert x-1\vert^{\nu_0}}
{1\over (x-1)}\int_1^xdy\ {(y+\beta^2)^{\nu_1-1}\over y^{\nu_0}}
\vert y-1\vert^{\nu_0}\cr
\noalign{\medskip}&&(A.4)\cr
I_2&=(1+\beta^2)\int_0^\infty    dx\ {x^{\nu_1}\over(x+\beta^2)^{\nu_0+1}}
{1\over\vert x-1\vert^{\nu_1}}
\int_1^xdy\ {(y+\beta^2)^{\nu_0}\over y^{\nu_1+1}}
{\vert y-1\vert^{\nu_1}\over y-1}\cr
}
$$
We have set $\sigma={2\phi_0^2\over n_0+n_1}$ and $\beta=\tan
2\alpha={C{E}\over C{\phi_0^2-E}}$.

According to (3.13) the Lyapunov exponent $\gamma(E)$ reads
$$
\gamma(E)=\phi_0\int_\alpha^{\pi+\alpha}d\t \cos 2\t
[P_0\T-P_1\T]
$$
{}From the expressions (A.1) and (A.2) for the stationary distributions
$P_0\T$ and $P_1\T$ in the interval $0<E<\phi_0^2$ we obtain after some
transformations
$$
\gamma(E)=N(E)[J_1-J_2]
\eqno(A.5)
$$
where
$$\eqalign{
J_1=\$&\int_0^\infty    dx
\left({x^2+\beta^2(2x-1)\over x^2+\beta^2}\right)
{1\over(x+\beta^2)^{\nu_1}}
{1\over\vert{1\over x}-1\vert^{\nu_0}}\cr
\noalign{\medskip}
&\int_1^xdy\ (y+\beta^2)^{\nu_1}
\left\vert{1\over y}-1\right\vert^{\nu_0}
\left[
{\nu_1\over x(x-1)}{1\over y+\beta^2}-{\nu_0\over x+\beta^2}{1\over y(y-1)}
\right]\cr}
\eqno(A.6)
$$
and $J_2$ is obtained from $J_1$ by the exchange between $\nu_1$ and $\nu_0$.

\medskip
\boxit{3pt}{b) Results for $E>\phi_0^2$}

 Here we use the parameter $ \ok {1\over\gamma}=C{E\over\phi_0^2}>1$ and the
function\break $\widetilde q\T=  \exp \bigg(
-\alpha_1\arctan({\tan\t+\gamma\over
C{1-\gamma^2}})-\alpha_0\arctan({\tan\t-\gamma\over
C{1-\gamma^2}}) \bigg)$ with
$\alpha_1= \ok {\gamma\over C{1-\gamma^2}}{n_1\over\phi_0}$ and
$\alpha_0= \ok {\gamma\over C{1-\gamma^2}}{n_0\over\phi_0}$ , and
 where arctan stands for the principal determination of $\tan^{-1}$
taking values in $\left[-{\pi\over 2},\ {\pi\over 2}\right]$.

\noindent The solutions $P_0\T$ and $P_1\T$ of (3.11) read for
$\t\in\left[-{\pi\over 2},\ {\pi\over 2}\right]$
$$\left\{\matrix{
P_0\T= {n_1\over \phi_0^2}N(E) \ok {\widetilde q\T\over {1\over\gamma}-\sin
2\t}
\left[
\int_0^\t{dt\over({1\over\gamma}+\sin 2t)\widetilde q(t)}+A
\right]\cr
\noalign{\bigskip}
P_1\T={n_0\over \phi_0^2}N(E) \ok {\widetilde q\T\over {1\over\gamma}+\sin 2\t}
\left[
\int_0^\t{dt\over({1\over\gamma}-\sin 2t)\widetilde q(t)}+B
\right]\cr
}\right.\eqno(A.7)
$$
where the constants A and B are choosen to ensure the $\pi$ periodicity of
$P_0\T$ and $P_1\T$
$$
A={1\over 1-e^{-(\alpha_0+\alpha_1)\pi}}
\int_{-{\pi\over 2}}^0{dt\over({1\over\gamma}+\sin 2t)\widetilde q(t)}
+{1\over e^{(\alpha_0+\alpha_1)\pi}-1}
\int_0^{\pi\over 2}{dt\over({1\over\gamma}+\sin 2t)\widetilde q(t)}
$$
and B is obtained from A by replacing $({1\over\gamma}+\sin 2t)$ by
$({1\over\gamma}-\sin 2t)$ in the integrals.

$N(E)$ is obtained by the normalization condition
$\int_{-{\pi\over 2}}^{\pi\over 2}d\t P_0\T={n_1\over n_0+n_1}$ and
$\gamma(E)$ by the expression
$$
\gamma(E)=\phi_0\int_{-{\pi\over 2}}^{\pi\over 2}d\t\cos 2\t \ \big(
P_0\T-P_1\T \big)
$$

Let us now consider the limit $E\to\phi_0^2$.
$N(E)$ is given by two very different expressions for $E<\phi_0^2$ and
$E>\phi_0^2$ but it is nevertheless continuous at the point $E=\phi_0^2$, as it
should by definition of the integrated density of states.
$N(\phi_0^2+0)=N(\phi_0^2-0)
={2\sigma\over I_1+I_2}$ where
$$\left\{\matrix{
I_1=\int_0^\infty     dx \ok \ e^{-{\mu_1\over x}+\mu_0x}
\int_x^\infty     {dy\over y}\ e^{{\mu_1\over y}-\mu_0y}\hfill\cr
\noalign{\medskip}
I_2=\int_0^\infty     dx \ok \ e^{{\mu_1\over x}-\mu_0x}
\int_0^x{dy\over y^2}\ e^{-{\mu_1\over y}+\mu_0y}\hfill\cr
}\right.
\eqno(A.8)
$$
However one should not expect to have more than continuity of $N(E)$ at this
point. In particular the derivative $\rho(E)={dN\over dE}$ has no reason to be
continuous at this point  (See V).
\def\arctg{\mathop{\rm arctg}\nolimits}
\def\oppropto{\mathop{\propto}}
\def\fio2{\phi_0^2}

\noindent{\bf APPENDIX B} : THE LYAPUNOV EXPONENT FOR $E<0$

 Here the dynamical equation for the phase $\t$ is
given in (2.21)
$$
{d\t\over dx}=-k\cos 2\t(x)-\phi(x)\sin 2\t(x)
\eqno(B.1)
$$
We have already seen in the discussion following (2.21) that $\t(x)$ is trapped
in an interval of the form $[-{\pi\over 2},\ 0]$ [modulo $\pi$].

 \noindent For the two step model
$\left(\matrix{
\phi_0>0\hfill\cr
\phi_1=-\phi_0\hfill\cr
}\right)$
defined in (III.1), a more accurate discussion shows that the phase $\t$ is
trapped in the interval $[\alpha_1,\ \alpha_0]$ [modulo $\pi$] where
$\alpha_0\in]-{\pi\over 4},\ 0[$ is the angle defined by
$$
\sin 2\alpha_0={-k\over C{k^2+\fio2}}
\eqno(B.2)
$$
and $\alpha_1=-{\pi\over 2}-\alpha_0$.

\noindent Indeed on an interval where $\phi(x)=\phi_0$ the phase $\t(x)$ is
attracted
towards $\alpha_0$ according to
${d\t\over dx}=-C{\fio2+k^2}\sin(2\t-2\alpha_0)$
and on the interval $\phi(x)=-\phi_0$, the phase $\t(x)$ is attracted towards
$\alpha_1$ according to
${d\t\over dx}=-C{\fio2+k^2}\sin(2\t-2\alpha_1)$

\noindent The stationary distributions $P_0\T$ and $P_1\T$ have therefore their
support on the
interval $[\alpha_1,\alpha_0]$.

 Using the relation (3.28), the system (3.27) can be rewritten as two distinct
equations for $P_0\T$ and $P_1\T$, whose solutions are for
$\t\in[\alpha_1,\alpha_0]$
$$\left\{\matrix{
P_0\T=A{q\T\over\sin(2\alpha_0-2\t)}\hfill\cr
\noalign{\medskip}
P_1\T=A{q\T\over\sin(2\t-2\alpha_1)}\hfill\cr
}\right.
\eqno(B.3)
$$
where
$$\left\{\matrix{
&q\T=\vert\tan(\t-\alpha_0)\vert^{\nu_0}\ \vert\tan(\t-\alpha_1)\vert^{\nu_1}
\hfill\cr
\noalign{\bigskip}\cr
&\nu_0={n_0\over 2C{\fio2+k^2}}\qquad \nu_1={n_1\over 2C{\fio2+k^2}}\hfill\cr
}\right.
\eqno(B.4)
$$
The constant A is determined by the normalization condition
$$
\int_{\alpha_1}^{\alpha_0} d\t\ P_0\T={n_1\over n_0+n_1}
\eqno(B.5)
$$

The expression (3.24) for the Lyapunov exponent
$\gamma(E)$
$$
\gamma(E)=\lim_{L\to \infty     }{1\over L}
\int_0^Ldx[-k\sin 2\t(x)+\phi(x)\cos 2\t(x)]
$$
can now be rewritten in terms of the stationary distributions $P_0\T$ and
$P_1\T$
$$\eqalign{
\gamma(E)&=\int_{\alpha_1}^{\alpha_0} d\t\ [-k\sin 2\t(P_0\T+P_1\T)+
\phi_0\cos 2\t(P_0\T-P_1\T)]\cr
&=C{\fio2+k^2}\int_{\alpha_1}^{\alpha_0} d\t\ [\cos(2\t-2\alpha_0)P_0\T+
\cos(2\t-2\alpha_1)P_1\T]\cr
}\eqno(B.6)
$$
\def\oppropto{\mathop{\propto}}
\def\arctg{\mathop{\rm arctg}\nolimits}
\def\fio2{\phi_0^2}

\noindent {\bf APPENDIX C} : PHASE EVOLUTION ON AN INTERVAL WHERE
$\phi(x)=\phi_0$

The integration of (5.1) requires to separate $k<\phi_0$ from $k>\phi_0$.

$\bullet$ For $k<\phi_0$ we introduce $\alpha_0\in\ ]0,{\pi\over 4}[$
satisfying
$\sin 2\alpha_0={k \over\phi_0}$.
$$\left\{\matrix{
{d\t\over dx}>0 \iff \t\ \hbox{[modulo}\ \pi]\ \in\ ]
-{\pi\over 2}-\alpha_0,\ \alpha_0\ [\hfill\cr
\ecart
{d\t\over dx}<0 \iff\t\ \hbox{[modulo}\ \pi]\ \in\ ]
\ \alpha_0,\ {\pi\over 2}-\alpha_0\ [\hfill\cr
}\right.
$$
Therefore $\vert\t_f-\t_i\vert$ cannot exceed $\pi$.\hfill\break
More precisely, if $(\t_i+\alpha_0)$ belongs to the interval
$]-{\pi\over 2}+n\pi,\ {\pi\over 2}+n\pi]$ with $n\in\ N$ then
$(\t_f+\alpha_0)$
stays within the same interval, and the corresponding tangents are related
through
$$
\tan(\t_f+\alpha_0)-\tan 2\alpha_0=[\tan(\t_i+\alpha_0)-\tan 2\alpha_0]
e^{-2\ell\sqrt{\fio2-E}}
\eqno(C.1)
$$

$\bullet$ For $k>\phi_0$ the increment $(\t_f-\t_i)$ is not limited.

\noindent The local primitive of the function involved in (5.1) reads
$$
\int^\phi{d\t\over
k-\phi_0\sin 2\t}
={1\over\sqrt{E-\fio2}}\arctan\left({\tan\phi-\gamma_0\over\sqrt{1-\gamma_0^2}}\right)
\quad\hbox{where}\quad\gamma_0={\phi_0\over k} $$ An increment of $\pi$ for the
phase $\t$ requires therefore the length $\tau$ given by
$$
\tau=\int_{-{\pi\over 2}}^{\pi\over 2}{d\t \over k-\phi_0\sin 2\t}
={\pi\over C{E-\fio2}}
\eqno(C.2)
$$
If we denote by $m\in N$ the integer part of $\left(\ell\over\tau\right)$, the
final phase reads
$$
\t_f=\t_i+m\pi+\varphi\ ,
\eqno(C.3)
$$
where the angle $\varphi\in\ [0,\pi\ [$ is determined by
$$
\ell=m\tau+{1\over\sqrt{E-\fio2}}
\left[\arctan\left(
{\tan\t-\gamma_0\over\sqrt{1-\gamma_0^2}}\right)
\right]_{\t_i}^{\t_i+\varphi}\ .
\eqno(C.4)
$$
Here $\arctan$ is a continuous determination of $\tan^{-1}$, and therefore
$$
\left[\arctan\left(
{\tan\t-\gamma_0\over\sqrt{1-\gamma_0^2}}\right)
\right]_{\t_i}^{\t_i+\varphi}\ \in\ [\ 0,\pi\ [\quad\hbox{for}\quad
\varphi\in\ [\ 0,\pi\ [\ .
$$
\noindent{\bf APPENDIX D} : PROBABILITY DENSITY ${\cal P}_L(x)$

 Let $\{\phi(x)\}$ be a random process which alternates in 2 states
$\phi(x)=\phi_0$ and\break $\phi(x)=-\phi_0$. We denote by $l_n\ (h_n)$
the successive sojourn lengths in state $\phi_0$ (resp. $-\phi_0$). We
are interested in the distribution of the random variable $A(L)$ which
measures the total length in state $\phi_0$ on the interval [0, L] (see for
instance [31]). The
total distance in the state $\phi(x)=-\phi_0$ on [0, L] is then simply
[L - A(L)].

The probability density ${\cal P}_L(x)$ such that
$$
{\cal P}_L(x)dx = Pr\{x<A(L)<x+dx\}
$$
naturally splits into four terms according to the values of $\phi$ at
the end points $\phi(0)=\pm\phi_0$ and $\phi(L)=\pm\phi_0$. In obvious
notations
one has
$$
{\cal P}_L(x)={\cal P}_L^{++}(x)+{\cal P}_L^{+-}(x)+{\cal P}_L^{-+}(x)+
{\cal P}_L^{--}(x)
\eqno(D1)
$$

Let us consider the case $\phi(0)=\phi_0,\ \phi(L)=\phi_0$. A typical
configuration of this type is represented in the following figure.
\vglue 4cm
\hglue 1cm\special{picture figa}
\medskip
\noindent The probability density ${\cal P}_L^{++}(x)$ therefore reads
$$
\eqalignno{
{\cal P}_L^{++}(x) dx &=Pr\big(\phi(0)=\phi_0\big) \times \cr
 &  \sum_{N=0}^\infty   Pr
\left(
 L-x -dx\leq \sum_{i=1}^{N}  h_i  \leq L-x
\right)
Pr
\left(
\sum_{j=1}^{N}\ell_j\leq x\leq\sum_{j=1}^{N+1}\ell_j
\right) \ \ \ \
&(D2) \cr
}$$
The probability involved in this expression can be obtained easily. We have
already seen that (3.3)
$$
Pr(\phi(0)=\phi_0)={n_1\over n_0+n_1}
$$
Each $\ell_i$ is distributed according to the probability density (3.1)
$$
f_0(\ell) = \theta (l) \  n_0 \ e^{-n_0\ell}
$$
whose Fourier transform reads
$$
\widehat f_0(p)\equiv\int_{-\infty }^{+\infty}dx\ f_0(x) \ e^{-ipx}={n_0\over
n_0+ip}
$$
The total length $\ y= \ok \sum_{i=1}^N\ell_i$ is a sum of independent
identically distributed random variables. It is distributed with the
probability density
$$
f_0^{\{N\}}(y)=\int_{-\infty}^{+\infty }{dp\over 2\pi}\ e^{ipy} \
\bigg[\widehat f_0(p) \bigg]^N =
\theta(y) \ n_0 \ {(n_0y)^{N-1}\over (N-1)!} \ e^{-n_0y}
\eqno(D3)
$$
The probability $Q_0^{\{N\}}(x)$ to have
${\displaystyle\left(
\sum_{i=1}^N\ell_i\leq x\leq \sum_{i=1}^{N+1}\ell_i
\right)}$
then reads
$$\eqalignno{
Q_0^{\{N\}}(x)&=Pr\left(\sum_{i=1}^N\ell_i\leq x\right)-
Pr\left(\sum_{i=1}^{N+1}\ell_i\leq x\right)=
\int_0^xdy
\left[f_0^{\{N\}}(y)-f_0^{\{N+1\}}(y)\right]\cr
&=\theta(x)\ {(n_0x)^N\over N!}\ e^{-n_0x}\ .&(D4)\cr
}$$
We also define $f_1^{\{N\}}(x)$ and $Q_1^{\{N\}}(x)$ by replacing $n_0$ by
$n_1$\hfill\break
$f_1^{\{N\}}(x)$ is the probability density to have
$\displaystyle\left(\sum_{i=1}^Nh_i= x\right)$\hfill\break
$Q_1^{\{N\}}(x)$ is the probability to have
$\displaystyle\left(\sum_{i=1}^Nh_i\leq x\leq\sum_{i=1}^{N+1}h_i\right)$

\noindent Eq. (D2) becomes
$$
{\cal P}_L^{++}(x)={n_1\over n_0+n_1}\sum_{N=0}^\infty  f_1^{\{N\}}(L-x)\
Q_0^{\{N\}}(x)
\eqno(D5)
$$

In a similar way we can compute ${\cal P}_L^{+-}(x)$ such that
$$
{\cal P}_L^{+-}(x)dx=Pr
\left\{x<A(L)<x+dx\ \vert\ \phi(0)=\phi_0,\ \phi(L)=-\phi_0\right\}
$$
by considering configurations represented in the following figure.
\vglue 4.5cm
\indent\special{picture figb}
\medskip

\noindent We get
$$\eqalignno{
{\cal P}_L^{+-}(x) dx&=Pr(\phi(0)=\phi_0)  \sum_{N=0}^\infty
Pr\left[x \leq \sum_{i=1}^{N+1}\ell_i \leq x +dx\right]
Pr\left[\sum_{i=1}^Nh_i\leq L-x\leq \sum_{i=1}^{N+1}h_i\right]\cr
&={n_1\over n_0+n_1}\sum_{N=0}^\infty  f_0^{\{N+1\}}(x) \
Q_1^{\{N\}}(L-x)&(D6)
}$$

We can now obtain ${\cal P}_L^{-+}(x)$ [resp. ${\cal P}_L^{--}(x)$] from
${\cal P}_L^{+-}(x)$ [resp. ${\cal P}_L^{++}(x)$] by the simple exchanges
$\left(\matrix{
n_0\longleftrightarrow n_1\hfill\cr
x\longleftrightarrow L-x\cr
}\right.$
so that the probability density ${\cal P}_L(x)$ (A1) finally reads
$$
\eqalignno{
{\cal P}_L(x)&={n_1\over n_0+n_1}\sum_{N=0}^\infty
\left[
f_1^{\{N\}}(L-x) \ Q_0^{\{N\}}(x)+f_0^{\{N+1\}}(x) \ Q_1^{\{N\}}(L-x)
\right]\cr
&+{n_0\over n_0+n_1}\sum_{N=0}^\infty
\left[
f_0^{\{N\}}(x) \ Q_1^{\{N\}}(L-x)+f_1^{\{N+1\}}(L-x) \ Q_0^{\{N\}}(x)
\right]&(D7)\cr
}$$
\centerline{\bf ACKNOWLEDGEMENTS}

\noindent We thank Jean-Philippe Bouchaud for many helpful discussions. We also
thank Frederic
Fougere for
his collaboration in the early stage of this work, and Olivier Martin for
reading the manuscript.
\Ref

\item{\hbox to\parindent{\enskip [1] \hfill}}
P.W. Anderson, {\it Phys. Rev.} {\bf 109} (1958) 1492.

\item{\hbox to\parindent{\enskip [2] \hfill}}
B. Souillard, "Les Houches", XLVI, J. Souletie, J. Vanimenus and R. Stora,
(1987).

\item{\hbox to\parindent{\enskip [3] \hfill}}
E. Witten, {\it Nucl. Phys.} {\bf B188} (1981) 513.

\item{\hbox to\parindent{\enskip [4] \hfill}}
R. Dutt, A. Khare and U.P. Sukhatme , {\it Am. J. Phys.} {\bf 59} (1991) 723.

\item{\hbox to\parindent{\enskip [5] \hfill}}
A. Comtet, A.D. Bandrauk and D.K. Campbell , {\it Phys. Lett.} {\bf 150B}
(1985) 159.

\item{\hbox to\parindent{\enskip [6] \hfill}}
A. Inomata and G. Junker , "Proceedings of International Symposium on Advanced
Topics in Quantum
Physics" (J.Q.Liang, M.L.Wang, S.N.Qiao and D.C.Su Eds.) Science Press,
Beijing, (1993) 61.

\item{\hbox to\parindent{\enskip [7] \hfill}}
N.G. Van Kampen, "Stochastic Processes in Physics and Chemistry", North
Holland,
Amsterdam (1981).

\item{\hbox to\parindent{\enskip [8] \hfill}}
J.P. Bouchaud, A. Comtet, A. Georges and P. Le Doussal , {\it Ann. Phys.} {\bf
201} (1990) 285.

\item{\hbox to\parindent{\enskip [9] \hfill}}
J.P. Bouchaud, A. Comtet, A. Georges and P. Le Doussal , {\it Europhysics.
Lett.} {\bf3} (6)
(1987) 653.

\item{\hbox to\parindent{\enskip [10] \hfill}}
B. Derrida and Y. Pomeau , {\it Phys. Rev .Lett.} {\bf 48} (1982) 627.

\item{\hbox to\parindent{\enskip [11] \hfill}}
J.P. Bouchaud and A. Georges,  {\it Phys. Rep.} {\bf 195} (1990) 127.

\item{\hbox to\parindent{\enskip [12] \hfill}}
G. Oshanin, S.F. Burlatsky, M. Moreau and B. Gaveau,  {\it Chem. Phys.} {\bf
178} (1993).

\item{\hbox to\parindent{\enskip [13] \hfill}}
C.W. Gardiner, "Handbook of Stochastic Methods", Springer Verlag,
New York (1982).

\item{\hbox to\parindent{\enskip [14] \hfill}}
D. Waxman and K.D. Ivanova-Moser, {\it Ann. Phys.} {\bf 226} (1993) 271.

\item{\hbox to\parindent{\enskip [15] \hfill}}
M.M. Benderskii and L.A. Pastur, {\it Sov. Phys. JETP} {\bf 30} (1970) 158.

\item{\hbox to\parindent{\enskip [16] \hfill}}
I.M. Lifshits, S.A. Gredeskul and L.A. Pastur, "Introduction to the theory of
disordered systems",
John Wiley and Sons, New York (1987).

\item{\hbox to\parindent{\enskip [17] \hfill}}
 L.A. Pastur and E.P. Feld'man, {\it Sov. Phys. JETP} {\bf 40} (1974) 241.

\item{\hbox to\parindent{\enskip [18] \hfill}}
P. Erdos and Z. Domanski, "Analogies in Optics and Micro Electronics" (W.van
Haeringen and
D.Lenstra Eds.) Kluwer Academic Publishers (1990) 49.

\item{\hbox to\parindent{\enskip [19] \hfill}}
G. Barton, A.J. Bray and A.J. McKane,  {\it Am. J. Phys.} {\bf 58} (1990) 751.

\item{\hbox to\parindent{\enskip [20] \hfill}}
E. Tossati, A. Vulpiani and M. Zannetti, {\it Physica A.} {\bf A164} (1990)
705.

\item{\hbox to\parindent{\enskip [21] \hfill}}
G. Theodorou and M.H. Cohen, {\it Phys. Rev. B} {\bf 13} (1976) 4597 ;
A. Bovier, {\it J.Stat.Phys.} {\bf 56} (1989) 645.

\item{\hbox to\parindent{\enskip [22] \hfill}}
J.P. Bouchaud, A. Comtet, A. Georges and P. Le Doussal , {\it J. Physique} {\bf
48} (1987) 1445.

\item{\hbox to\parindent{\enskip [23] \hfill}}
B.V. Gnedenko and A.N. Kolmogorov, "Limit Distributions for Sums of Independent
variables",
Addison-Wesley, Reading, MA, 1954.

\item{\hbox to\parindent{\enskip [24] \hfill}}
A. Comtet, A. Georges and P. Le Doussal, {\it Phys. Lett.} {\bf B208} (1988)
487.

\item{\hbox to\parindent{\enskip [25] \hfill}}
A.V. Smilga, {\it Phys.Rev.} {\bf D46} (1992) 5598.

\item{\hbox to\parindent{\enskip [26] \hfill}}
H. Takayama,Y.R. Lin-Liu and K. Maki,  {\it Phys. Rev. B} {\bf 21} (1980) 2388.

\item{\hbox to\parindent{\enskip [27] \hfill}}
D.K. Campbell and A.R. Bishop, {\it Nucl. Phys.} {\bf B200} (1982) 297.

\item{\hbox to\parindent{\enskip [28] \hfill}}
B.C. Xu and S.E. Trullinger,  {\it Phys. Rev. Lett.} {\bf 57} (1986) 3113.

\item{\hbox to\parindent{\enskip [29] \hfill}}
K.A. Loparo and X. Feng,  {\it Siam J. Appl. Math} {\bf 53} (1993) 283.

\item{\hbox to\parindent{\enskip [30] \hfill}}
C. Aslangul, N. Potier, P. Chvosta and D. Saint-James,  {\it Phys.Rev.} {\bf
E47} (1993) 1610.

\item{\hbox to\parindent{\enskip [31] \hfill}}
S. Karlin and H.M. Taylor, "A first course in stochastic processes", Academic
Press, New-York
(1975)
\vfill\eject
\def\fio2{\phi_0^2}
\Fig
\fig1
The random force $\phi(x)$ takes alternatively two values $\phi_0$
and $\phi_1$ on intervals whose lengths $\ell_i's$ and $h_i's$ are independent
random variables,
distributed according to (3.1) unless otherwise stated.

\fig2
The Schr\"odinger potential $(\phi^2+\phi')$  as a function
of x  in the case $\phi_0=-\phi_1$. Delta functions appear each time $\phi(x)$
jumps.

\fig3
Computer simulations of the average density of states $N(E)$ for
$\phi_0=-\phi_1=1$ and $ n_0=n_1 =n $, with respectively
{\bf a)} $n= 5.$ {\bf b)} $n=1.$ {\bf c)} $n= 0.5$ {\bf d)} $n= 0.1$
 $N(E)$ appears to be an
increasing function of n. In general, its derivative is not a continuous
function of the energy; a discontinuity can occur at $E=\fio2$.

\fig4
The $\pi$-periodic stationary distributions  $W_0\T$ and $W_1\T$
 are plotted, respectively in full curve and in dashed curve,
for the case $\phi_0=-\phi_1=1,\ E=0.4,\ n_0=n_1=2$.
 $W_0$ and $W_1\T$ defined just before (5.5) are the same distributions
as $P_0\T$ and $P_1\T$ defined just before (3.8) up to a normalization constant
(5.8).
 Note the peak of $W_0\T$ at $\t=\alpha_0\approx 0.34$, and the peak of $W_1\T$
at
 $\t=\alpha_1=\alpha_0 - {\pi \over 2}$.
 For further explanations, see text.

\fig5
The same as Fig. 4 except for $ n_0=n_1=4$ instead of $n_0=n_1=2$.
The lengths of the intervals on which $\phi(x)$ is constant are shorter, and
 the peaks at $\alpha_0$ and $\alpha_1$ are therefore attenuated.

\fig6
Lyapunov exponents $\gamma(E)$ are plotted for $\phi_0=-\phi_1=1$ and
 $ n_0=n_1 =n $, with respectively
{\bf a)} $n= 1.$\  {\bf b)} $n=2.$\  {\bf c)} $n= 4.$ \ \ \ \
For these three cases, $\gamma(E)$ vanishes at zero energy, but very slowly
(3.17).

\fig7
The same as Fig. 6 except for $n_0=2n_1 =n$ and $\gamma(0)=\vert F_0\vert=1/3$,
according to (2.10).

\fig8
The density of states $N(E)$, for the case $\phi_0=1$ and $\phi_1=-C{2}$,
 when the steps lengths are drawn according to the
same broad distribution, behaving as (5.10) with $\alpha=0.5$, and whose first
moment therfore diverges.
 Curves a) b) c) represent
three different simulation events. Obviously, $N(E)$ is no longer
self-averaging (5.19).
\eject
\end